# Incomplete Information in RDF

Charalampos Nikolaou
University of Athens
University Campus, Ilisia
15784 Athens, Greece
charnik@di.uoa.gr

Manolis Koubarakis
University of Athens
University Campus, Ilisia
15784 Athens, Greece
koubarak@di.uoa.gr

## ABSTRACT

We extend RDF with the ability to represent property values that exist, but are unknown or partially known, using constraints. Following ideas from the incomplete information literature, we develop a semantics for this extension of RDF, called RDF$^i$, and study SPARQL query evaluation in this framework.

## 1. INTRODUCTION

Incomplete information has been studied in-depth in relational databases [12, 6] and knowledge representation. It is also an important issue in Semantic Web frameworks such as RDF, description logics, and OWL 2 especially given that all these systems rely on the Open World Assumption (OWA). Making the OWA means that we cannot capture negative information implicitly, i.e., if a formula $\phi$ is not entailed by our knowledge base, we cannot assume its negation as in the Closed World Assumption (CWA).

Application knowledge captured by databases and knowledge bases is often incomplete, thus the OWA is a useful assumption to make. In general, the richer an application domain is, the more possible it is that a framework based on incomplete information will be required. Incomplete information can also arise even if we start from complete databases, e.g., in relational view updates, data integration, data exchange, etc., thus the detailed study of incomplete information has been a recurring theme in the literature throughout the years.

In the context of the Web, incomplete information has recently been studied in detail for XML [2, 5]. As Semantic Web technologies achieve maturity and gain acceptance in a wide variety of application domains through the creation of ontologies and linked data pools, we expect the study of issues related to incomplete information to gain more attention in the Semantic Web community as well. There have been some recent papers that confirm our expectations.

[9] introduces the concept of *anonymous timestamps* in general temporal RDF graphs, i.e., graphs containing quads of the form $(s,p,o)[t]$ where $t$ is a timestamp (a natural number) or an anonymous timestamp $x$ stating that the triple $(s,p,o)$ is valid in some unknown time point $x$. [11] subsequently extends the concept of general temporal RDF graphs of [9] so that one is allowed to express temporal constraints involving anonymous timestamps using a formula $\phi$ which is a conjunction of order constraints $x_1 \ OP \ x_2$ where $OP$ is an arithmetic comparison operator such as $<, \leq$, etc. [11] calls $c$-temporal graphs the resulting pairs $(G, \phi)$ where $G$ is a general temporal RDF graph and $\phi$ is a conjunction of constraints. [11] defines a semantics for $c$-temporal graphs and studies the relevant problem of entailment.

More recently, [4] examines the question of whether SPARQL is an appropriate language for RDF given the OWA typically associated with the framework. It defines a *certain answer* semantics for SPARQL query evaluation based on well-known ideas from incomplete information research. According to this semantics, if $G$ is an RDF graph then evaluating a SPARQL query $q$ over $G$ is defined as evaluating $q$ over all graphs $H \supseteq G$ that are possible extensions of $G$ according to the OWA, and then taking the intersection of all answers. [4] shows that if we evaluate a monotone graph pattern (e.g., one using only the operators AND, UNION, and FILTER) using the well-known W3C semantics, we get the same result we would get if we used the certain answer semantics. The converse also holds, thus monotone SPARQL graph patterns are exactly the ones that have this nice property. However, OPTIONAL (OPT) is not a monotone operator and the two semantics do not coincide for it. [4] defines the notion of weak monotonicity that appears to capture the intuition behind OPT, and shows that a SPARQL query $q$ is weakly monotone if and only if evaluating $q$ under the W3C semantics gives the same result as evaluating $q$ under a new semantics appropriate for weakly monotone queries. Finally, [4] shows that the fragment of SPARQL consisting of the well-designed graph patterns defined originally in [26] is weakly monotone.

### 1.1 Contributions

In this paper we continue the line of research started by [9, 11, 4] and study in a general way an important kind of incomplete information that has so far been ignored in the context of RDF. Our contributions are the following.

First, we extend RDF with the ability to define a new kind of literals for each datatype. These literals will be called *e-literals* ("e" comes from the word "existential") and can be used to represent values of properties that *exist but are unknown or partially known*. Such information is abundant in recent applications where RDF is being used (e.g., sensor networks, the modeling of geospatial information, etc.). In the proposed extension of RDF, called RDF[i] (where "i" stands for "incomplete"), e-literals are allowed to appear only in the object position of triples.

Previous research on incomplete information in databases and knowledge representation has shown that in many applications, having the ability to state *constraints* about values that are only partially known is a very desirable feature and leads to the development of very expressive formalisms [6, 15]. In the spirit of this tradition, RDF[i] allows partial information regarding property values represented by e-literals to be expressed by a quantifier-free formula of a first-order *constraint language* $\mathcal{L}$. Thus, RDF[i] extends the concept of an RDF graph to the concept of an RDF[i] *database* which is a pair $(G, \phi)$ where $G$ is an RDF graph possibly containing triples with e-literals in their object positions, and $\phi$ is a quantifier-free formula of $\mathcal{L}$. Our recent workshop paper [22] motivates the need for introducing RDF[i] by concentrating on the representation of incomplete spatial knowledge.

Following ideas from the incomplete information literature [12, 6], we develop a semantics for RDF[i] databases and SPARQL query evaluation. The semantics defines the set of possible RDF graphs corresponding to an RDF[i] database and the fundamental concept of certain answer for SPARQL query evaluation over an RDF[i] database. We transfer the well-known concept of *representation system* from [12] to the case of RDF[i], and show that CONSTRUCT queries without blank nodes in their templates and using only operators AND, UNION, and FILTER or the restricted fragment of graph patterns corresponding to the well-designed patterns of [4] can be used to define a representation system for RDF[i]. Our results for the monotonicity of CONSTRUCT queries (even in the case of well-designed patterns that contain operator OPT) indicate their importance and sets an interesting subject to explore in theoretical treatments of RDF.

We define the fundamental concept of certain answer to SPARQL queries over RDF[i] databases and present an algorithm for its computation. Finally, we present preliminary complexity results for computing certain answers by considering equality, temporal, and spatial constraint languages $\mathcal{L}$ and the class of CONSTRUCT queries of our representation system. Our results show that the data complexity of evaluating a query of this class over RDF[i] databases increases from LOGSPACE (the upper bound for evaluating queries from this class over RDF graphs [26]) to coNP-complete for the case of equality and temporal constraints. This result is in line with similar complexity results for querying incomplete information in relational databases [6, 14]. The same coNP-completeness bound is shown for the case of spatial constraints on rectangles in $\mathbb{Q}^2$ [14]. For topological constraints over more general spatial regions (regular, closed subsets of $\mathbb{Q}^2$), the best upper bound that we can show is EXPTIME. To the best of our knowledge, it is an open problem how to achieve better complexity results in this case. The complexity of the closely related problem of SPARQL query evaluation over RDF graphs (e.g., as manifested in geospatial extensions stSPARQL [18] and GeoSPARQL [23]) has not been investigated so far in any detail, and it remains an open problem as well.

The organization of the paper is as follows. Section 2 presents the properties that we expect constraint languages to have so that they can be used in RDF[i]. In addition, it defines some useful constraint languages that will be used in the paper. Section 3 introduces RDF[i] and then Section 4 defines its semantics. Section 5 defines the evaluation of SPARQL queries over RDF[i] databases. Section 6 presents fragments of SPARQL that can be used to define a representation system for RDF[i]. Section 7 gives an algorithm for computing the certain answer for SPARQL queries over RDF[i] databases and presents our complexity results. Sections 8 and 9 discuss related and future work respectively. Last, the Appendix contains the complete proofs of the results established in this paper.

## 2. CONSTRAINT LANGUAGES

We will consider many-sorted first-order languages, structures, and theories. Every language $\mathcal{L}$ will be interpreted over a *fixed* structure, called the *intended structure*, which will be denoted by $\mathbf{M}_\mathcal{L}$. If $\mathbf{M}_\mathcal{L}$ is a structure then $Th(\mathbf{M}_\mathcal{L})$ will denote the theory of $\mathbf{M}_\mathcal{L}$, i.e., the set of sentences of $\mathcal{L}$ that are true in $\mathbf{M}_\mathcal{L}$. For every language $\mathcal{L}$, we will distinguish a class of quantifier free formulae called $\mathcal{L}$-*constraints*. The atomic formulae of $\mathcal{L}$ will be included in the class of $\mathcal{L}$-constraints. There will also be two distinguished $\mathcal{L}$-constraints *true* and *false* with obvious semantics.

Every first-order language $\mathcal{L}$ we consider has a distinguished equality predicate, denoted by EQ, with the standard semantics. The class of $\mathcal{L}$-constraints is assumed to: a) contain all formulae $t_1$ EQ $t_2$ where $t_1, t_2$ are terms of $\mathcal{L}$, and b) be *weakly closed under negation*, i.e., the negation of every $\mathcal{L}$-constraint is equivalent to a disjunction of $\mathcal{L}$-constraints. This property is needed in Section 7 where the certain answer to a SPARQL query over an RDF[i] database is computed.

Section A of the Appendix defines formally various constraint languages that allow us to explore the scope of modeling possibilities that RDF[i] offers. These languages are ECL, diPCL, dePCL, PCL, TCL and RCL. ECL is the first order language of equality constraints of the form $x$ EQ $y$ and $x$ EQ $c$ (where $x, y$ are variables and $c$ is a constant) interpreted over an infinite domain [6]. When used in RDF[i], this language allows us to extend RDF with the ability to represent "marked nulls" as in classical relational databases [12]. The languages diPCL and dePCL are the first order languages of temporal difference constraints of the form $x - y \leq c$ interpreted over the integers (diPCL) or the rationals (dePCL) [14]. These are constraint languages that allow RDF[i] to represent incomplete temporal information as in [9, 11] and older works such as [16]. PCL, TCL and RCL are spatial constraint languages and are defined in detail below. PCL is the language that we have used in our introductory paper [22] and we will also use it in the examples of this paper. PCL, TCL and RCL will be referred to in Section 8 where the power of RDF[i] for geospatial modeling is compared with other modeling frameworks.

### 2.1 The Languages PCL, TCL, and RCL

The language *PCL* (*P*olygon *C*onstraint *L*anguage) allows us to represent topological properties of non-empty, regular closed subsets of $\mathbb{Q}^2$ (we call them *regions*). PCL is a first-order language with the following 6 binary predicate symbols corresponding to the topological relations of RCC-8 calculus [28]: DC, EC, PO, EQ, TPP, and NTPP. The constant symbols of PCL represent polygons in $\mathbb{Q}^2$. We will write these constants as conjunctions of linear constraints in quotes (half-space representation of the convex polygon). The terms and atomic formulae of PCL are defined as follows. Constants and variables are *terms*. An *atomic formula* of PCL (PCL-constraint) is a formula of the form $t_1 \; R \; t_2$ where $t_1, t_2$ are terms and $R$ is one of the above predicates.

The intended structure for PCL, denoted by $\mathbf{M}_{PCL}$, has the set of non-empty, regular closed subsets of $\mathbb{Q}^2$ as its domain. $\mathbf{M}_{PCL}$ interprets each constant symbol by the corresponding polygon in $\mathbb{Q}^2$ and each of the predicate symbols by the corresponding topological relation of RCC-8 [28].

Language TCL (*T*opological *C*onstraint *L*anguage) is defined like PCL, but now terms can only be variables (no topological reasoning with constants, i.e., landmarks [20], is allowed). Language RCL (*R*ectangle *C*onstraint *L*anguage) is a simpler first-order constraint language that represents information about *rectangles* in $\mathbb{Q}^2$ using rational constants and order or difference constraints ($x-y \leq c$) on the vertices of rectangles. RCL has essentially the same expressive power with dePCL, but it's been carefully crafted for rectangles.

## 3. THE RDF$^i$ FRAMEWORK

As in theoretical treatments of RDF [26], we assume the existence of pairwise-disjoint, countably infinite sets $I$, $B$, and $L$ that contain IRIs, blank nodes, and literals respectively. We also assume the existence of a datatype map $M$ [10] and distinguish a set of datatypes $A$ from $M$ for which e-literals are allowed. Finally, we assume the existence of a many-sorted first order constraint language $\mathcal{L}$ with the properties discussed in Section 2. $\mathcal{L}$ is related to the datatype map $M$ in the following way:

- The set of sorts of $\mathcal{L}$ is the set of datatypes $A$ of $M$.

- The set of constants of $\mathcal{L}$ is the union of the lexical spaces of the datatypes in $A$.

- $\mathbf{M}_\mathcal{L}$ interprets every constant $c$ of $\mathcal{L}$ with sort $d$ by its corresponding value given by the lexical-to-value mapping of the datatype $d$ in $A$.

The set of constants of $\mathcal{L}$ (equivalently: the set of literals of the datatypes in $A$) will be denoted by $C$. We also assume the existence of a countably infinite set of e-literals for each datatype in $A$ and use $U$ to denote the union of these sets. By convention, the identifiers of e-literals will start with an underscore, e.g., _R5. $C$ and $U$ are assumed to be disjoint from each other and from $I$, $B$, and $L$. The set of RDF$^i$ *terms*, denoted by $T$, can now be defined as the union $I \cup B \cup L \cup C \cup U$.

In the rest of our examples we will assume that $\mathcal{L}$ is PCL, so $C$ is the set of all polygons in $\mathbb{Q}^2$ written in the linear constraint syntax of Section 2.

We now define the basic concepts of RDF$^i$: e-triples, conditional triples, conditional graphs, global constraints, and databases. Triples in RDF$^i$ (called e-triples) are as in RDF but now e-literals are also allowed in the object position. Combining an e-triple with a conjunction of $\mathcal{L}$-constraints, we get a conditional triple. Graphs in RDF$^i$ are conditional and consist of sets of conditional triples. Global constraints are simply Boolean combinations of $\mathcal{L}$-constraints. The combination of a conditional graph and a global constraint is called a database.

DEFINITION 3.1. *An* e-triple *is an element of the set* $(I \cup B) \times I \times T$. *If* $(s, p, o)$ *is an e-triple, $s$ will be called the subject, $p$ the predicate, and $o$ the object of the triple. A* conditional triple *is a pair* $(t, \theta)$ *where $t$ is an e-triple and $\theta$ is a conjunction of $\mathcal{L}$-constraints. If $(t, \theta)$ is a conditional triple, $\theta$ will be called the condition of the triple.*

DEFINITION 3.2. *A* global constraint *is a Boolean combination of $\mathcal{L}$-constraints.*

DEFINITION 3.3. *A* conditional graph *is a set of conditional triples. An* RDF$^i$ database *$D$ is a pair $D = (G, \phi)$ where $G$ is a conditional graph and $\phi$ a global constraint.*

In the rest of the paper, when we want to refer to standard RDF constructs we will write "RDF triple" and "RDF graph" so that no confusion with RDF$^i$ is possible.

EXAMPLE 3.4. *The following pair is an* RDF$^i$ *database.*

( { ((hotspot1, type, Hotspot), true),
    ((fire1, type, Fire), true),
    ((hotspot1, correspondsTo, fire1), true),
    ((fire1, occuredIn, _R1), true) },

  _R1 NTPP "x $\geq$ 6 $\wedge$ x $\leq$ 23 $\wedge$ y $\geq$ 8 $\wedge$ y $\leq$ 19" )

EXAMPLE 3.5. *The following pair is an* RDF$^i$ *database with a disjunctive global constraint.*

( { ((hotspot1, type, Hotspot), true),
    ((fire1, type, Fire), true),
    ((hotspot1, correspondsTo, fire1), true),
    ((fire1, occuredIn, _R1), true),
    ((fire2, occuredIn,
              "x $\geq$ 6 $\wedge$ x $\leq$ 23 $\wedge$ y $\geq$ 8 $\wedge$ y $\leq$ 19"), true) },

(_R1 NTPP "x $\geq$ 6 $\wedge$ x $\leq$ 23 $\wedge$ y $\geq$ 8 $\wedge$ y $\leq$ 19" $\wedge$
 _R1 NTPP "x $\geq$ 10 $\wedge$ x $\leq$ 21 $\wedge$ y $\geq$ 12 $\wedge$ y $\leq$ 17") $\vee$
          _R1 PO "x $\geq$ 2 $\wedge$ x $\leq$ 6 $\wedge$ y $\geq$ 4 $\wedge$ y $\leq$ 8" )

## 4. SEMANTICS OF RDF$^i$

The semantics of RDF$^i$ are inspired by [12]. An RDF$^i$ database $D = (G, \phi)$ corresponds to a set of possible RDF graphs each one representing a possible state of the real world. This set of possible graphs captures completely the semantics of an RDF$^i$ database. The global constraint $\phi$ determines the number of possible RDF graphs corresponding to $D$; there is one RDF graph for each solution of $\phi$ obtained by considering the e-literals of $\phi$ as variables and solving the constraint $\phi$.

EXAMPLE 4.1. *Let $D = (G, \phi)$ be the RDF$^i$ database given in Example 3.4. Database $D$ mentions a hotspot, which is located in a region that is inside but does not intersect with the boundary of rectangle defined by the points $(6, 8)$ and $(23, 19)$. The same knowledge can be represented by an (infinite) set of possible RDF graphs, one for each rectangle inside $P$. Two of these graphs are:*

$G_1 = $ { (hotspot1, type, Hotspot), (fire1, type, Fire),
    (hotspot1, correspondsTo, fire1),
    (fire1, occuredIn, "x $\geq$ 11 $\wedge$ x $\leq$ 15 $\wedge$ y $\geq$ 13 $\wedge$ y $\leq$ 15")}

$G_2 = $ { (hotspot1, type, Hotspot), (fire1, type, Fire),
    (hotspot1, correspondsTo, fire1),
    (fire1, occuredIn, "x $\geq$ 10 $\wedge$ x $\leq$ 21 $\wedge$ y $\geq$ 12 $\wedge$ y $\leq$ 17")}

In order to be able to go from RDF$^i$ databases to the equivalent set of possible RDF graphs, the notion of *valuation* is needed. Informally, a valuation maps an e-literal to a specific constant from $C$.

DEFINITION 4.2. *A valuation $v$ is a function from $U$ to $C$ assigning to each e-literal from $U$ a constant from $C$.*

We denote by $v(t)$ the application of valuation $v$ to an e-triple $t$. $v(t)$ is obtained from $t$ by replacing any e-literal $\_l$ appearing in $t$ by $v(\_l)$ and leaving all other terms the same. If $\theta$ is a formula of $\mathcal{L}$ (e.g., the condition of a conditional triple or the global constraint of a database) then $v(\theta)$ denotes the application of $v$ to formula $\theta$. The expression $v(\theta)$ is obtained from $\theta$ by replacing all e-literals $\_l$ of $\theta$ by $v(\_l)$.

Next, we give the definition of applying a valuation to a conditional graph.

DEFINITION 4.3. *Let $G$ be a conditional graph and $v$ a valuation. Then $v(G)$ denotes the RDF graph*

$$\{v(t) \mid (t, \theta) \in G \text{ and } \mathbf{M}_\mathcal{L} \models v(\theta)\}.$$

The set of valuations that satisfy the global constraint of an RDF$^i$ database determines the set of possible RDF graphs that correspond to it. This set of graphs is denoted using the function *Rep* as it is traditional in incomplete relational databases.

DEFINITION 4.4. *Let $D = (G, \phi)$ be an RDF$^i$ database. The set of RDF graphs corresponding to $D$ is the following:*

$$Rep(D) = \{H \mid \text{there exists a valuation } v$$
$$\text{such that } \mathbf{M}_\mathcal{L} \models v(\phi) \text{ and } H \supseteq v(G)\}$$

In incomplete relational databases [12], *Rep* is a *semantic* function: it is used to map a table (a syntactic construct) to a set of relational instances (i.e., a set of possible words, a semantic construct). According to the well-known distinction between model theoretic and proof theoretic approaches to relational databases, *Rep* and the approaches based on it [12, 6] belong to the model theoretic camp. However, the use of function *Rep* in the above definition is different. *Rep* takes an RDF$^i$ database (a syntactic construct) and maps it to a set of possible RDF graphs (a syntactic construct again). This set of possible graphs can then be mapped to a set of possible worlds using the well-known RDF model theory [10]. This is a deliberate choice in our work since we want to explore which well-known tools from incomplete relational databases carry over to the RDF framework.

Notice that the definition of *Rep* above uses the containment relation instead of equality. The reason for this is to capture the OWA that the RDF model makes. By using the containment relation, $Rep(D)$ includes all graphs $H$ containing at least the triples of $v(G)$. In this respect, we follow the approach of [4, Section 3], where the question of whether SPARQL is a good language for RDF is examined in the light of the fact that RDF adopts the OWA. To account for this, an RDF graph $G$ is seen to correspond to a set of possible RDF graphs $H$ such that $G \subseteq H$ (in the sense of the OWA: all triples in $G$ also hold in $H$). The above definition takes this concept of [4] to its rightful destination: the full treatment of incomplete information in RDF. As we have already noted in the introduction, the kinds of incomplete information we study here for RDF has not been studied in [4]; only the issue of OWA has been explored there.

The following notation will be useful below.

NOTATION 1. *Let $\mathcal{G}$ be a set of RDF graphs and $q$ a SPARQL query. The expression $\bigcap \mathcal{G}$ will denote the set $\bigcap_{G \in \mathcal{G}} G$. The expression $[\![q]\!]_\mathcal{G}$, which extends the notation of [26] to the case of sets of RDF graphs, will denote the element-wise evaluation of $q$ over $\mathcal{G}$, that is,*

$$[\![q]\!]_\mathcal{G} = \{[\![q]\!]_G \mid G \in \mathcal{G}\}.$$

Given the semantics of an RDF$^i$ database as a set of possible RDF graphs, what is an appropriate definition for the answer to a certainty query? This is captured by the following definition of *certain answer* which extends the corresponding definition of Section 3.1 of [4] by applying it to a more general incomplete information setting.

DEFINITION 4.5. *Let $q$ be a query and $\mathcal{G}$ a set of RDF graphs. The* certain answer *to $q$ over $\mathcal{G}$ is the set $\bigcap [\![q]\!]_\mathcal{G}$.*

EXAMPLE 4.6. *Let us consider the following query over the database of Example 3.4: "Find all fires that have occurred in a region which is a non-tangential proper part of the rectangle defined by the points $(2, 4)$ and $(28, 22)$". The certain answer to this query is the set of mappings $\{\{?F \to \text{fire1}\}\}$.*

## 5. EVALUATING SPARQL ON RDF$^i$ DATABASES

Let us now discuss how to evaluate SPARQL queries on RDF$^i$ databases. We will use the algebraic syntax of SPARQL presented in [17]. We will consider only the monotone graph pattern fragment of SPARQL which uses only the AND, UNION, and FILTER operators [4]. We will deal with both SELECT and CONSTRUCT query forms. Due to the presence of e-literals, query evaluation now becomes more complicated and is similar to query evaluation for conditional tables [12, 6]. The exact details will be given later in this section.

We use set semantics for query evaluation by extending the SPARQL query evaluation approach of [26]. Blank nodes are interpreted as in SPARQL, i.e., as constants different from each other. Notice that this is not the same as the semantics of blank nodes in RDF model theory [10] where they are treated as existentially quantified variables.

We assume the existence of the following disjoint sets of variables: (i) the set of *normal query variables* $V_n$ that range over IRIs, blank nodes, or RDF literals, and (ii) the set of *special query variables* $V_s$ that range over literals from the set $C$ or e-literals from the set $U$. We use $V$ to denote the set of all variables $V_n \cup V_s$. Set $V$ is assumed to be disjoint from the set of terms $T$ we defined in Section 3.

We first define the concept of *e-mapping* ("*e*" from the word "existential") which extends the concept of mapping of [17] with the ability to have an e-literal as value of a special query variable.

DEFINITION 5.1. *An e-mapping $\nu$ is a partial function $\nu : V \to T$ such that $\nu(x) \in I \cup B \cup L$ if $x \in V_n$ and $\nu(x) \in C \cup U$ if $x \in V_s$.*

EXAMPLE 5.2. *The following are e-mappings.*

$\mu_1 = \{\ ?F \to \text{fire1}, ?S \to \text{"}x \geq 1 \land x \leq 2 \land y \geq 1 \land y \leq 2\text{"} \ \}$
$\mu_2 = \{\ ?F \to \text{fire1}, ?S \to \_R1 \ \}$
$\mu_3 = \{\ ?F \to \text{fire1}, ?S \to \_R2 \ \}$
$\mu_4 = \{\ ?F \to \text{fire1} \ \}$

The notions of domain and restriction of an *e*-mapping as well as the notion of compatibility of two *e*-mappings are defined as for mappings in the obvious way [26] (we also use the same notation for them).

We now extend the concept of *e*-mapping and define conditional mappings, i.e., mappings that are equipped with a condition which constrains e-literals that appear in the *e*-mapping.

DEFINITION 5.3. *A* conditional mapping $\mu$ *is a pair* $(\nu, \theta)$ *where $\nu$ is an e-mapping and $\theta$ is a conjunction of $\mathcal{L}$-constraints.*

EXAMPLE 5.4. *The following are conditional mappings.*

$\mu_1 = (\{?F \to \text{fire1}, ?S \to \text{"x} \geq 1 \land \text{x} \leq 2 \land \text{y} \geq 1 \land \text{y} \leq 2\text{"}\}, \text{true})$
$\mu_2 = (\{?F \to \text{fire1}, ?S \to \_R1\},$
$\qquad \_R1 \text{ NTPP "x} \geq 0 \land \text{x} \leq 10 \land \text{y} \geq 0 \land \text{y} \leq 10\text{"})$
$\mu_3 = (\{?F \to \text{fire1}, ?S \to \_R1\},$
$\qquad (\_R1 \text{ NTPP } \_R2) \land$
$\qquad (\_R2 \text{ DC "x} \geq 0 \land \text{x} \leq 1 \land \text{y} \geq 0 \land \text{y} \leq 1\text{"}) )$
$\mu_4 = (\{?F \to \text{fire1}, ?S \to \_R1\}, \text{true})$

Notice that conditional mappings with constraint *true*, such as $\mu_4$ above, are logically equivalent to *e*-mappings.

The notions of domain and restriction for a conditional mapping are now defined as follows.

DEFINITION 5.5. *The* domain *of a conditional mapping* $\mu = (\nu, \theta)$, *denoted by* $dom(\mu)$, *is the domain of $\nu$, i.e., the subset of $V$ where the partial function $\nu$ is defined.*

DEFINITION 5.6. *Let* $\mu = (\nu, \theta)$ *be a conditional mapping with domain $S$ and $W \subseteq S$. The* restriction *of the mapping $\mu$ to $W$, denoted by $\mu_{|W}$, is the mapping $(\nu_{|W}, \theta)$ where $\nu_{|W}$ is the restriction of mapping $\nu$ to $W$.*

We now define the basic notion of triple pattern.

DEFINITION 5.7. *A* triple pattern *is an element of the set* $(I \cup V) \times (I \cup V) \times (I \cup L \cup C \cup U \cup V)$.

Note that we do not allow blank nodes to appear in a triple pattern as in standard SPARQL since such blank nodes can equivalently be substituted by new query variables.

If $p$ is a triple pattern, $var(p)$ denotes the variables appearing in $p$. A conditional mapping can be applied to a triple pattern. Let $\mu = (\nu, \theta)$ be a conditional mapping and $p$ a triple pattern such that $var(p) \subseteq dom(\mu)$. We denote by $\mu(p)$ the triple obtained from $p$ by replacing each variable $x \in var(p)$ by $\nu(x)$.

We now introduce the notion of compatible conditional mappings as in [26].

DEFINITION 5.8. *Two conditional mappings* $\mu_1 = (\nu_1, \theta_1)$ *and* $\mu_2 = (\nu_2, \theta_2)$ *are* compatible *if the e-mappings* $\nu_1$ *and* $\nu_2$ *are compatible, i.e., for all* $x \in dom(\mu_1) \cap dom(\mu_2)$, *we have* $\nu_1(x) = \nu_2(x)$.

EXAMPLE 5.9. *Mappings* $\mu_1$ *and* $\mu_2$ *from Example 5.4 are not compatible, while mappings* $\mu_2$ *and* $\mu_3$ *are.*

To take into account e-literals, we also need to define another notion of compatibility of two conditional mappings.

DEFINITION 5.10. *Two conditional mappings* $\mu_1 = (\nu_1, \theta_1)$ *and* $\mu_2 = (\nu_2, \theta_2)$ *are* possibly compatible *if for all* $x \in dom(\mu_1) \cap dom(\mu_2)$, *we have* $\nu_1(x) = \nu_2(x)$ *or at least one of* $\nu_1(x), \nu_2(x)$ *where* $x \in V_s$ *is an e-literal from* $U$.

EXAMPLE 5.11. *Conditional mappings* $\mu_1$, $\mu_2$, *and* $\mu_3$ *from Example 5.4 are pairwise possibly compatible.*

If two conditional mappings are possibly compatible, then we can define their join as follows.

DEFINITION 5.12. *Let* $\mu_1 = (\nu_1, \theta_1)$ *and* $\mu_2 = (\nu_2, \theta_2)$ *be possibly compatible conditional mappings. The* join $\mu_1 \bowtie \mu_2$ *is a new conditional mapping* $(\nu_3, \theta_3)$ *where:*

i. $\nu_3(x) = \nu_1(x) = \nu_2(x)$ *for each* $x \in dom(\mu_1) \cap dom(\mu_2)$ *such that* $\nu_1(x) = \nu_2(x)$.

ii. $\nu_3(x) = \nu_1(x)$ *for each* $x \in dom(\mu_1) \cap dom(\mu_2)$ *such that* $\nu_1(x)$ *is an e-literal and* $\nu_2(x)$ *is a literal from* $C$.

iii. $\nu_3(x) = \nu_2(x)$ *for each* $x \in dom(\mu_1) \cap dom(\mu_2)$ *such that* $\nu_2(x)$ *is an e-literal and* $\nu_1(x)$ *is a literal from* $C$.

iv. $\nu_3(x) = \nu_1(x)$ *for* $x \in dom(\mu_1) \cap dom(\mu_2)$ *such that both* $\nu_1(x)$ *and* $\nu_2(x)$ *are e-literals.*

v. $\nu_3(x) = \nu_1(x)$ *for* $x \in dom(\mu_1) \setminus dom(\mu_2)$.

vi. $\nu_3(x) = \nu_2(x)$ *for* $x \in dom(\mu_2) \setminus dom(\mu_1)$.

vii. $\theta_3$ *is* $\theta_1 \land \theta_2 \land \xi_1 \land \xi_2 \land \xi_3$ *where:*

- $\xi_1$ *is* $\bigwedge_i \_v_i$ EQ $\_t_i$, *where the* $\_v_i$*'s and* $\_t_i$*'s are all the pairs of e-literals* $\nu_1(x)$ *and* $\nu_2(x)$ *from Case (iv) above. If there are no such pairs, then* $\xi_1$ *is true.*

- $\xi_2$ *is* $\bigwedge_i \_w_i$ EQ $l_i$ *where the* $\_w_i$*'s and* $l_i$*'s are all the pairs of e-literals* $\nu_1(x)$ *and literals* $\nu_2(x)$ *from the set $C$ from Case (ii) above. If there are no such pairs, then* $\xi_2$ *is true.*

- $\xi_3$ *is* $\bigwedge_i \_w_i$ EQ $l_i$ *where the* $\_w_i$*'s and* $l_i$*'s are all the pairs of e-literals* $\nu_2(x)$ *and literals* $\nu_1(x)$ *from the set $C$ from Case (iii) above. If there are no such pairs, then* $\xi_3$ *is true.*

The predicate EQ used in the above definition is the equality predicate of $\mathcal{L}$.

EXAMPLE 5.13. *If* $\mu_1$ *and* $\mu_2$ *are the conditional mappings of Example 5.4, then:*

$\mu_1 \bowtie \mu_2 = (\{?F \to \text{fire1}, ?S \to \_R1\}, \text{ true } \land$
$\qquad \_R1 \text{ EQ "x} \geq 1 \land \text{x} \leq 2 \land \text{y} \geq 1 \land \text{y} \leq 2\text{"} \land$
$\qquad \_R1 \text{ NTPP "x} \geq 0 \land \text{x} \leq 10 \land \text{y} \geq 0 \land \text{y} \leq 10\text{"} )$

For two sets of conditional mappings $\Omega_1$ and $\Omega_2$, the operation of join is now defined as follows.

$\Omega_1 \bowtie \Omega_2 = \{\mu_1 \bowtie \mu_2 \mid \mu_1 \in \Omega_1, \mu_2 \in \Omega_2 \text{ are possibly}$
$\qquad\qquad\qquad \text{compatible conditional mappings}\}$

The reader is invited to compare this definition with the definition of join of mappings for RDF [26]. The new thing with conditional mappings is that due to the presence of e-literals, we have to anticipate the possibility that two mappings from $\Omega_1$ and $\Omega_2$ become compatible when e-literals are substituted by constants from $C$. We anticipate this case by adding relevant constraints to the condition of a mapping.

The operation of union is defined as in the standard case:

$$\Omega_1 \cup \Omega_2 = \{\mu \mid \mu \in \Omega_1 \text{ or } \mu \in \Omega_2\}$$

We now define the operator of difference:

$\Omega_1 \setminus \Omega_2 =$
$\{\mu_1 \in \Omega_1 \mid \text{for all } \mu_2 \in \Omega_2, \mu_1 \text{ and } \mu_2 \text{ are not compatible}\} \cup$
$\{(\nu, \theta') \mid \mu = (\nu, \theta) \in \Omega_1 \text{ and } \mu_1 = (\nu_1, \theta_1), \ldots, \mu_n = (\nu_n, \theta_n)$
$\in \Omega_2 \text{ such that } \mu_i, \mu \text{ are possibly compatible for all}$
$1 \leq i \leq n, \text{ and for every other } \mu_j \in \Omega_2 \text{ different than}$
the $\mu_i$'s, $\mu_j$ and $\mu$ are not compatible. In this case, $\theta'$ is

$$\theta \bigwedge_i \left( \theta_i \supset \bigvee_x \left( \neg(\mu(x) \text{ EQ } \mu_i(x)) \right) \right)$$

for every $x \in dom(\mu) \cap dom(\mu_i) \cap V_s$ and $1 \leq i \leq n\}$

The reader is invited to compare this definition with the definition of difference in [26]. The new thing in RDF$^i$ is that we have to anticipate the possibility that a mapping $\mu$ from $\Omega_1$ is not compatible with all the mappings of $\Omega_2$ (i.e., it should be included in the difference) due to the presence of e-literals in it given some constraints. These constraints are added to the condition of $\mu$.

EXAMPLE 5.14. *Let* $\Omega_1 = \{\mu_{11}, \mu_{12}\}$, $\Omega_2 = \{\mu_{21}, \mu_{22}\}$ *be sets of conditional mappings such that*

$\mu_{11} = (\{?F \to \text{fire1}, ?S \to \_R1\},$
    $\_R1 \text{ NTPP } "x \geq 0 \land x \leq 10 \land y \geq 0 \land y \leq 10")$

$\mu_{12} = (\{?F \to \text{fire1}, ?S \to "x \geq 1 \land x \leq 2 \land y \geq 1 \land y \leq 2"\},$
    $\_R1 \text{ PO } "x \geq 0 \land x \leq 10 \land y \geq 0 \land y \leq 10")$

$\mu_{21} = (\{?F \to \text{fire2}\}, \text{ true})$

$\mu_{22} = (\{?F \to \text{fire1}, ?S \to "x \geq 1 \land x \leq 2 \land y \geq 1 \land y \leq 2"\}, \text{true})$

*Then,* $\Omega_1 \setminus \Omega_2 = \{\mu\}$ *where* $\mu$ *has been constructed from* $\mu_{11} \in \Omega_1$ *and is the following mapping:*

$\mu = (\{?F \to \text{fire1}, ?S \to \_R1\},$
    $(\_R1 \text{ NTPP } "x \geq 0 \land x \leq 10 \land y \geq 0 \land y \leq 10") \land$
    $\neg(\_R1 \text{ EQ } "x \geq 1 \land x \leq 2 \land y \geq 1 \land y \leq 2") )$

The operation of left-outer join is defined as in the standard case:

$$\Omega_1 \bowtie\!\!\!\!\!\!\!\!\!\!\!\!\!\!\!{\scriptstyle \supset} \Omega_2 = (\Omega_1 \bowtie \Omega_2) \cup (\Omega_1 \setminus \Omega_2)$$

It has been noted in [26] that the OPT operator of SPARQL (the counterpart of the left-outer join algebraic operator) can be used to express difference in SPARQL. For data models that make the OWA, such an operator is unnatural since negative information cannot be expressed. However, we deliberately include operator OPT because if it is combined with operators AND and FILTER under certain syntactic restrictions, it turns out that the resulting graph patterns cannot express a difference operator anymore [4]. In particular, the class of graph patterns produced by this syntactic restriction are known as *well-designed graph patterns*. Well-designed graph patterns are discussed in more depth in Section 6 where representation systems for RDF$^i$ are investigated.

We can now define the result of evaluating a graph pattern over an RDF$^i$ database (the definition of graph patterns is omitted). Given the previous operations on sets of mappings, graph pattern evaluation in RDF$^i$ can now be defined exactly as in standard SPARQL for RDF graphs [26] except for the case of evaluating a triple pattern.

DEFINITION 5.15. *Let* $D = (G, \phi)$ *be an* RDF$^i$ *database. Evaluating a graph pattern $P$ over database $D$ is denoted by* $[\![P]\!]_D$ *and is defined recursively as follows:*

1. *If $P$ is the triple pattern $(s, p, o)$ then we have two cases. If $o$ is a literal from the set $C$ then*

    $[\![P]\!]_D = \{\mu = (\nu, \theta) \mid dom(\mu) = var(P) \text{ and}$
    $(\mu(P), \theta) \in G\} \cup$
    $\{\mu = (\nu, (\_l \text{ EQ } o) \land \theta) \mid dom(\mu) = var(P),$
    $((\nu(s), \nu(p), \_l), \theta) \in G \text{ and } \_l \in U\}$

    *else*

    $[\![P]\!]_D = \{\mu = (\nu, \theta) \mid dom(\mu) = var(P), (\mu(P), \theta) \in G\}$

2. *If $P$ is $P_1$ AND $P_2$ then $[\![P]\!]_D = [\![P_1]\!]_D \bowtie [\![P_2]\!]_D$.*

3. *If $P$ is $P_1$ UNION $P_2$ then $[\![P]\!]_D = [\![P_1]\!]_D \cup [\![P_2]\!]_D$.*

4. *If $P$ is $P_1$ OPT $P_2$ then $[\![P]\!]_D = [\![P_1]\!]_D \bowtie\!\!\!\!\!\!\!\!\!\!\!\!\!\!\!{\scriptstyle \supset} [\![P_2]\!]_D$.*

In the first item of the above definition the "else" part is to accommodate the case in which evaluation can be done as in standard SPARQL. This is the case in which the object part of the triple pattern is not a literal from $C$. The "if" part accommodates the case in which the triple pattern involves a literal $o$ from the set $C$. Here, there are two alternatives: the graph contains a conditional triple matching with every component of the triple pattern (i.e., a triple which has $o$ in the object position) or it contains a conditional triple with an e-literal $\_l$ from $U$ in the object position. We catch a possible match for the second case by adding in the condition of the mapping the constraint that restricts the value of e-literal $\_l$ to be equal to the literal $o$ of the triple pattern (i.e., the constraint $\_l$ EQ $o$). In all cases of the first item of the above definition, since the triples in the database are conditional, their conditions become parts of the conditions of the mappings in the answer.

EXAMPLE 5.16. *Let us first give two examples for the evaluation of triple patterns over the database $D$ of Example 3.5.*

$[\![(?F, \text{occuredIn}, "x \geq 1 \land x \leq 2 \land y \geq 1 \land y \leq 2")]\!]_D = \{\ \} \cup$
$\{(\{?F \to \text{fire1}\}, \_R1 \text{ EQ } "x \geq 1 \land x \leq 2 \land y \geq 1 \land y \leq 2")\}$

$[\![(?F, \text{occuredIn}, "x \geq 6 \land x \leq 23 \land y \geq 8 \land y \leq 19")]\!]_D =$
$\{(\{?F \to \text{fire2}\}, \text{true})\} \cup \{(\{?F \to \text{fire1}\},$
    $\_R1 \text{ EQ } "x \geq 6 \land x \leq 23 \land y \geq 8 \land y \leq 19")\}$

*These examples correspond to the "if" part of the first item of Definition 5.15 in which the triple pattern involves a literal from the set $C$.*

EXAMPLE 5.17. *Let us now give an example of an evaluation of graph pattern $P_1$ AND $P_2$ over the database $D$ of Example 3.4, where $P_1, P_2$ are the triple patterns $(?F, \text{type}, \text{Fire})$ and $(?F, \text{occuredIn}, ?R)$ respectively. According to the above definition, we have:*

$[\![(P_1 \text{ AND } P_2)]\!]_D = [\![P_1]\!]_D \bowtie [\![P_2]\!]_D =$
$[\![(?F, \text{type}, \text{Fire})]\!]_D \bowtie [\![(?F, \text{occuredIn}, ?R)]\!]_D =$
$\{(\{?F \to \text{fire1}\}, \text{true})\} \bowtie \{(\{?F \to \text{fire1}, ?R \to \_R1\}, \text{true})\} =$
    $\{(\{?F \to \text{fire1}, ?R \to \_R1\}, \text{true})\}$

*The evaluation of both triple patterns $P_1, P_2$ corresponds to the "else" part of the first item of Definition 5.15. In this case evaluation is done as in standard SPARQL, but here conditions of matched triples have to be transferred to the respective answer, i.e., we have conditional mappings.*

Let us now consider the operator FILTER. It is natural to allow FILTER graph patterns to contain conjunctions of $\mathcal{L}$-constraints as expressions that constrain query variables, e.g., constraints like ?X NTPP ?Y or ?X EQ "x ≥ 1 ∧ x ≤ 2 ∧ y ≥ 1 ∧ y ≤ 2" when $\mathcal{L}$ is PCL as in our examples.

The evaluation of FILTER graph patterns involving $\mathcal{L}$-constraints can now be defined as follows. Notice that the evaluation does not check for satisfaction of the constraints as in standard SPARQL [26], but simply imposes these constraints on the mappings that are in the answer of the graph pattern involved.

DEFINITION 5.18. *Given an* $\text{RDF}^i$ *database* $D = (G, \phi)$, *a graph pattern P and a conjunction of* $\mathcal{L}$-*constraints R, we have:*

$$[\![P \; FILTER \; R]\!]_D = \{\mu' = (\nu, \theta') \mid \mu = (\nu, \theta) \in [\![P]\!]_D$$
$$\text{and } \theta' \text{ is } \theta \wedge \nu(R) \}$$

In the above definition, $\nu(R)$ denotes the application of $e$-mapping $\nu$ to condition $R$, i.e., the conjunction of $\mathcal{L}$-constraints obtained from $R$ when each variable $x$ of $R$ which also belongs to $dom(\nu)$ is substituted by $\nu(x)$.

The extension of FILTER to the case that $R$ is a Boolean combination of $\mathcal{L}$-constraints is now easy to define and is omitted. Similarly, the extension of FILTER to the case that $R$ contains also other built-in conditions of standard SPARQL [26] is easy to define and is omitted as well.

The following example illustrates the definition and shows that the purpose of constraint $\nu(R)$ is to deal in a uniform way with the case that the object of a triple is a constant from $C$ or an e-literal from $U$. Notice that $\nu(R)$ is required because mappings in our case can contain variables with e-literals as values, thus we might not be able to deduce their satisfaction yet. Thus, evaluation of FILTERs is "lazy". In an implementation, one can also simplify constraints at this stage; such issues are beyond the scope of this paper.

EXAMPLE 5.19. *Based on the evaluation of the graph pattern of Example 5.17, the evaluation of the graph pattern* $((P_1 \; AND \; P_2) \; FILTER \; R)$, *where R is the PCL-constraint* (?R NTPP "x ≥ 10 ∧ x ≤ 21 ∧ y ≥ 12 ∧ y ≤ 17"), *is the following:*

$[\![(P_1 \; AND \; P_2) \; FILTER \; R]\!]_D =$
$[\![((?F, type, Fire) \; AND \; (?F, occuredIn, ?R)) \; FILTER$
(?R NTPP "x ≥ 10 ∧ x ≤ 21 ∧ y ≥ 12 ∧ y ≤ 17")$]\!]_D =$
$\{(\{?F \to \text{fire1}, ?R \to \_R1\},$
    $\_R1 \; NTPP \; "x ≥ 10 ∧ x ≤ 21 ∧ y ≥ 12 ∧ y ≤ 17")\}$

The next definition defines the concept of a SELECT query [26].

DEFINITION 5.20. *A* SELECT *query is a pair* $(W, P)$ *where W is a set of variables from the set V and P is a graph pattern.*

EXAMPLE 5.21. *Let us consider the following query over the database of Example 3.4:* "Find all fires that have occurred in a region which is a non-tangential proper part of rectangle defined by the points (10, 12) and (21, 17)". *This query can be expressed as follows:*

$(\{?F\}, (?F, type, Fire) \; AND \; (?F, occuredIn, ?R)$
    FILTER (?R NTPP "x ≥ 10 ∧ x ≤ 21 ∧ y ≥ 12 ∧ y ≤ 17"))

The next definition defines the notion of answer to a SELECT query. In contrast to SELECT queries over RDF graphs, SELECT queries over $\text{RDF}^i$ databases have answers that consist of conditional mappings so they might be harder to understand.

DEFINITION 5.22. *Let* $q = (W, P)$ *be a* SELECT *query. The answer to q over an* $\text{RDF}^i$ *database* $D = (G, \phi)$ *(in symbols* $[\![q]\!]_D$) *is the set of conditional mappings* $\{\mu_{|W} \mid \mu \in [\![P]\!]_D\}$.

The conditional mappings of the answer to a query might contain e-literals. These literals are constrained by the global constraint $\phi$, therefore $\phi$ can be understood to be implicitly included in the answer (this can also be done formally by considering answers to be pairs).

EXAMPLE 5.23. *The answer to the query from Example 5.21 can be obtained from the evaluation of the respective graph pattern from Example 5.19. The answer is a set that contains only the following mapping:*

$(\{?F \to \text{fire1}\}, \_R1 \; NTPP \; "x ≥ 10 ∧ x ≤ 21 ∧ y ≥ 12 ∧ y ≤ 17")$

*This answer is conditional. Because the information in the database of Example 3.4 is indefinite (the exact geometry of* `_R1` *is not known), we cannot say for sure whether* `fire1` *satisfies the requirements of the query. These requirements are satisfied under the condition given in the above mapping.*

Let us now introduce the notion of a *template* and define the CONSTRUCT query form.

DEFINITION 5.24. *A template E is a finite subset of the set* $(T \cup V) \times (I \cup V) \times (T \cup V)$.

Thus, the elements of a template are like triple patterns but blank nodes are also allowed in the subject and object positions. We denote by $var(E)$ and $blank(E)$ the set of variables and set of blank nodes appearing in the elements of $E$ respectively.

DEFINITION 5.25. *A* CONSTRUCT *query is a pair* $(E, P)$ *where E is a template and P a graph pattern.*

EXAMPLE 5.26. *Let us consider the query of Example 5.21. A new version of this query using the* CONSTRUCT *query form is:*

$(\{(?F, type, Fire)\}, (?F, type, Fire) \; AND \; (?F, occuredIn, ?R)$
    FILTER (?R NTPP "x ≥ 10 ∧ x ≤ 21 ∧ y ≥ 12 ∧ y ≤ 17"))

Next we define what it means for a conditional mapping to be applied to a template.

DEFINITION 5.27. *Let* $\mu = (\nu, \theta)$ *be a conditional mapping and E a template. We denote by* $\mu(E)$ *the application of conditional mapping* $\mu$ *to template E.* $\mu(E)$ *is obtained from E by replacing in E every variable x of* $var(E) \cap dom(\mu)$ *by* $\nu(x)$.

Templates are used to specify the graph that results from the evaluation of a CONSTRUCT query.

EXAMPLE 5.28. *Let us consider the template* $E = \{(?F, type, ?Z), (?F, occuredIn, ?S)\}$ *and mapping* $\mu_4$ *from Example 5.4. The result of applying* $\mu_4$ *to E is the following set:*

$\{(\text{fire1}, type, ?Z), (\text{fire1}, occuredIn, \_R1)\}$

*Notice that Definition 5.27 does not require a conditional mapping to share any variables with the template to which it is applied. As a consequence, the first element of $\mu_4(E)$ is not a valid e-triple, i.e., it is not an element of the set $(I \cup B) \times I \times T$. Such a triple is dropped from the answer to a* CONSTRUCT *query (see Definition 5.30 below).*

Next we define the concept of answer to a CONSTRUCT query. The definition extends the specification of standard SPARQL [27] to account for the RDF$^i$ framework and follows the formal approach of [25]. Before we give the definition, we need to introduce the notion of *renaming function*.

DEFINITION 5.29. *Let $E$ be a template, $P$ a graph pattern, and $D = (G, \phi)$ an RDF$^i$ database. The set $\{f_\mu \mid \mu \in [\![P]\!]_D\}$ is a set of* renaming functions *for $E$ and $[\![P]\!]_D$ if the following properties are satisfied: 1) the domain of every function $f_\mu$ is $blank(E)$ and its range is a subset of $(B \setminus blank(G))$, 2) every function $f_\mu$ is one-to-one, and 3) for every pair of distinct mappings $\mu_1, \mu_2 \in [\![P]\!]_D$, $f_{\mu_1}, f_{\mu_2}$ have disjoint ranges.*

The application of a renaming function $f_\mu$ to a template $E$ is denoted by $f_\mu(E)$ and results in renaming the blank nodes of $E$ according to $f_\mu$.

DEFINITION 5.30. *Let $q = (E, P)$ be a* CONSTRUCT *query, $D = (G, \phi)$ an RDF$^i$ database and $F = \{f_\mu \mid \mu \in [\![P]\!]_D\}$ a fixed set of renaming functions. The* answer *to $q$ over $D$ (in symbols $[\![q]\!]_D$) is the RDF$^i$ database $D' = (G', \phi)$ where*

$$G' = \bigcup_{\mu = (\nu, \theta) \in [\![P]\!]_D} \{(t, \theta) \mid t \in (\mu(f_\mu(E)) \cap ((I \cup B) \times I \times T))\}.$$

In the above definition, renaming functions are used to ensure that brand new blank nodes are created for each conditional mapping $\mu$. The intersection with $(I \cup B) \times I \times T$ makes sure that no illegal triples are returned as answers (see Example 5.28 above).

EXAMPLE 5.31. *The answer to the* CONSTRUCT *query from Example 5.26 can be obtained from the evaluation of the respective graph pattern from Example 5.19. The answer is the following RDF$^i$ database:*

( { ((fire1, type, Fire),
    _R1 NTPP "x ≥ 10 ∧ x ≤ 21 ∧ y ≥ 12 ∧ y ≤ 17") },

 _R1 NTPP "x ≥ 6 ∧ x ≤ 23 ∧ y ≥ 8 ∧ y ≤ 19" )

## 6. REPRESENTATION SYSTEMS FOR RDF$^i$

Let us now recall the semantics of RDF$^i$ as given by *Rep*. $Rep(D)$ is the set of possible RDF graphs corresponding to an RDF$^i$ database $D$. Clearly, if we were to evaluate a query $q$ over $D$, we could use the semantics of RDF$^i$ and evaluate $q$ over any RDF graph of $Rep(D)$ as follows:

$$[\![q]\!]_{Rep(D)} = \{[\![q]\!]_G \mid G \in Rep(D)\}$$

However, this is not the best answer we wish to have in terms of representation; we queried an RDF$^i$ database and got an answer which is a set of RDF graphs. Any well-defined query language should have the *closure* property, i.e., the output (answer) should be of the same type as the input. Ideally, we would like to have an RDF$^i$ database as the output. Thus, we are interested in finding an RDF$^i$ database $[\![q]\!]_D$ representing the answer $[\![q]\!]_{Rep(D)}$. This requirement is translated to the following formula:

$$Rep([\![q]\!]_D) = [\![q]\!]_{Rep(D)} \quad (1)$$

Formula (1) allows us to compute the answer to any query over an RDF$^i$ database in a consistent way with respect to the semantics of RDF$^i$ without having the need to apply the query on all possible RDF graphs. $[\![q]\!]_D$ can be computed using the algebra of Section 5 above. But can the algebra of Section 5 compute always such a database $[\![q]\!]_D$ representing $[\![q]\!]_{Rep(D)}$? In other words, can we prove (1) for all SPARQL queries considered in Section 5? The answer is *no* in general. The following example modelled after [7] illustrates this negative fact.

EXAMPLE 6.1. *Consider the RDF$^i$ database $D = (G, \phi)$, where $G = \{((s, p, o), true)\}$ and $\phi = true$, i.e., $D$ contains the single triple $(s, p, o)$ where $s, p, o \in I$. Consider now a* CONSTRUCT *query $q$ over $D$ that selects all triples having $s$ as the subject. The algebraic version of query $q$ would be $(\{(s, ?p, ?o)\}, (s, ?p, ?o))$ and evaluated as $[\![q]\!]_D$ using Definition 5.30. Then, the triple $(s, p, o)$ and nothing else is in the resulting database $[\![q]\!]_D$. However, equation (1) is not satisfied, since for instance $(c, d, e)$ occurs in some $g \in Rep([\![q]\!]_D)$ according to the definition of Rep, whereas $(c, d, e) \notin g$ for all $g \in [\![q]\!]_{Rep(D)}$.*

Note that the above counterexample to (1) exploits only the fact that RDF makes the OWA. In other words, the counterexample would hold for any approach to incomplete information in RDF which respects the OWA. Thus, unless the CWA is adopted, which we do not want to do since we are in the realm of RDF, condition (1) has to be relaxed[1].

In the rest of this section we follow the literature of incomplete information [12, 6] and show how (1) can be weakened. The key concept for achieving this is the concept of certain answer we defined earlier. Given a fixed fragment of SPARQL $\mathcal{Q}$, two RDF$^i$ databases cannot be distinguished by $\mathcal{Q}$ if they give the same certain answer to every query in $\mathcal{Q}$. The next definition formalizes this fact using the concept of $\mathcal{Q}$-equivalence. Originally this concept was defined for incomplete relational databases in [12].

DEFINITION 6.2. *Let $\mathcal{Q}$ be a fragment of SPARQL, and $\mathcal{G}$, $\mathcal{H}$ two sets of RDF graphs. $\mathcal{G}$ and $\mathcal{H}$ are called $\mathcal{Q}$-equivalent (denoted by $\mathcal{G} \equiv_\mathcal{Q} \mathcal{H}$) if they give the same certain answer to every query in the language, that is, $\bigcap [\![q]\!]_\mathcal{G} = \bigcap [\![q]\!]_\mathcal{H}$ for all $q \in \mathcal{Q}$.*

We can now define the notion of a *representation system* which gives a formal characterization of the correctness of computing the answer to a query directly on an RDF$^i$ database instead of using the set of possible graphs given by *Rep*. The definition of representation system (originally defined in [12] for incomplete relational databases) corresponds to the notion of *weak query system* defined in the same context by [6].

DEFINITION 6.3. *Let $\mathcal{D}$ be the set of all RDF$^i$ databases, $\mathcal{G}$ the set of all RDF graphs, $Rep : \mathcal{D} \to \mathcal{G}$ a function determining the set of possible RDF graphs corresponding to an*

---
[1]If the CWA is adopted, we can prove (1) using similar techniques to the ones that enable us to prove Theorem 6.14 below.

RDF$^i$ *database, and $\mathcal{Q}$ a fragment of SPARQL. The triple $\langle \mathcal{D}, Rep, \mathcal{Q} \rangle$ is a* representation system *if for all $D \in \mathcal{D}$ and all $q \in \mathcal{Q}$, there exists an* RDF$^i$ *database $[\![q]\!]_D \in \mathcal{D}$ such that the following condition is satisfied:*

$$Rep([\![q]\!]_D) \equiv_\mathcal{Q} [\![q]\!]_{Rep(D)}$$

The next step towards the development of a representation system for RDF$^i$ and SPARQL is to introduce various fragments of SPARQL that we will consider and define the notions of *monotonicity* and *coinitiality* as is done in [12]. As in Section 5, our only addition to standard SPARQL is the extension of FILTERs with another kind of conditions that are constraints of $\mathcal{L}$. We also consider the fragment of SPARQL graph patterns known as *well-designed*. Well-designed graph patterns form a practical fragment of SPARQL graph patterns that include the OPT operator and it has been showed in [26, 4] that that they have nice properties, such as lower combined complexity than in the general case, a normal form which is useful for optimization, and they are also weakly monotone. Thus, it is worth studying them in the context of RDF$^i$. Section B of the Appendix contains formal definitions and relevant background results for well-designed graph patterns.

NOTATION 2. *We denote by $\mathcal{Q}_\mathcal{F}^C$ (resp. $\mathcal{Q}_\mathcal{F}^S$) the set of all* CONSTRUCT *(resp.* SELECT*) queries consisting of triple patterns, and graph pattern expressions from class $\mathcal{F}$. We also denote by $\mathcal{Q}_{WD}^C$ (resp. $\mathcal{Q}_{WD}^S$) the set of all* CONSTRUCT *(resp.* SELECT*) queries consisting of well-designed graph patterns. Last, we denote by $\mathcal{Q}_\mathcal{F}^{C'}$ all* CONSTRUCT *queries without blank nodes in their templates.*

The following definition introduces the concept of monotone fragments of SPARQL applied to RDF graphs. Then, Proposition 6.5 give us some fragments of SPARQL that are monotone.

DEFINITION 6.4. *A fragment $\mathcal{Q}$ of SPARQL is* monotone *if for every $q \in \mathcal{Q}$ and RDF graphs $G$ and $H$ such that $G \subseteq H$, it is $[\![q]\!]_G \subseteq [\![q]\!]_H$.*

PROPOSITION 6.5. *The following results hold with respect to the monotonicity of SPARQL: a) Language $\mathcal{Q}_{AUF}^S$ is monotone. b) The presence of OPT or* CONSTRUCT *makes a fragment of SPARQL not monotone. c) Language $\mathcal{Q}_{AUF}^{C'}$ is monotone. d) Language $\mathcal{Q}_{WD}^{C'}$ is monotone.*

Parts $a) - c)$ of the above proposition are trivial extensions of relevant results in [4]. However, part $d)$ is an interesting result showing that the weak monotonicity property of well-designed graph patterns suffices to get a monotone fragment of SPARQL containing the OPT operator, i.e., the class of CONSTRUCT queries without blank nodes in their templates. This is a result that cannot be established for the case of SELECT queries and with this respect CONSTRUCT queries deserve closer attention.

Monotonicity is a sufficient property for establishing our results about representation systems. Thus, in the following, we focus on the monotone query languages $\mathcal{Q}_{AUF}^{C'}$ and $\mathcal{Q}_{WD}^{C'}$.

DEFINITION 6.6. *Let $\mathcal{G}$ and $\mathcal{H}$ be sets of RDF graphs. We say that $\mathcal{G}$ and $\mathcal{H}$ are* coinitial*, denoted by $\mathcal{G} \approx \mathcal{H}$, if for any $G \in \mathcal{G}$ there exists $H \in \mathcal{H}$ such that $H \subseteq G$, and for any $H \in \mathcal{H}$ there exists $G \in \mathcal{G}$ such that $G \subseteq H$.*

EXAMPLE 6.7. *The following sets are coinitial.*

$\mathcal{G} = \{\{(a,b,c),(a,e,d),(a,f,g)\},\{(a,b,c),(a,e,d)\},\{(a,b,c)\}\}$
$\mathcal{H} = \{\{(a,b,c),(a,e,d)\},\{(a,b,c)\}\}$

A direct consequence of the definition of coinitial sets is that they have the same greatest lower-bound elements with respect to the subset relation. In the above example, the greatest lower bound is $\bigcap \mathcal{G} = \bigcap \mathcal{H} = \{(a,b,c)\}$.

PROPOSITION 6.8. *Let $\mathcal{Q}$ be a monotone fragment of SPARQL and $\mathcal{G}$ and $\mathcal{H}$ sets of RDF graphs. If $\mathcal{G} \approx \mathcal{H}$ then, for any $q \in \mathcal{Q}$, it holds that $[\![q]\!]_\mathcal{G} \approx [\![q]\!]_\mathcal{H}$.*

LEMMA 6.9. *Let $\mathcal{G}$ and $\mathcal{H}$ be sets of RDF graphs. If $\mathcal{G}$ and $\mathcal{H}$ are coinitial then $\mathcal{G} \equiv_{\mathcal{Q}_{AUF}^{C'}} \mathcal{H}$.*

We will now present our main theorem which characterizes the evaluation of monotone $\mathcal{Q}_{AUF}^{C'}$ and $\mathcal{Q}_{WD}^{C'}$ queries (Theorem 6.14). Before we do this, we need a few definitions and preliminary results. The first definition allows us to apply a valuation to a conditional mapping. By applying a valuation to a conditional mapping, we get an *ordinary mapping* like in the case of RDF simply by disregarding the constraint that results since it is equivalent to *true*.

DEFINITION 6.10. *Let $v : U \to C$ be a valuation and $\mu = (\nu, \theta)$ a conditional mapping such that $\mathbf{M}_\mathcal{L} \models v(\theta)$. Then $v(\mu)$ denotes the mapping that results from substituting in e-mapping $\nu$ the constant $v(\_l)$ for each e-literal $\_l$.*

In a similar way, we can extend a valuation $v$ to a set of mappings $\Omega$ as follows.

DEFINITION 6.11. *Let $\Omega$ be a set of conditional mappings and $v : U \to C$ a valuation. Then*

$$v(\Omega) = \{v(\mu) \mid \mu = (\nu, \theta) \in \Omega \text{ and } \mathbf{M}_\mathcal{L} \models v(\theta)\}.$$

The next definition allows us to apply a valuation to an RDF$^i$ database.

DEFINITION 6.12. *Let $v : U \to C$ be a valuation and $D = (G, \phi)$ an* RDF$^i$ *database such that $\mathbf{M}_\mathcal{L} \models v(\phi)$. Then $v(D)$ denotes the RDF graph $v(G)$.*

PROPOSITION 6.13. *Let $D = (G, \phi)$ be an* RDF$^i$ *database, $q$ a query from a monotone fragment $\mathcal{Q}$ of SPARQL, and $v$ a valuation such that $\mathbf{M}_\mathcal{L} \models v(\phi)$. Then, $v([\![q]\!]_D) = [\![q]\!]_{v(D)}$ implies $Rep([\![q]\!]_D) \approx [\![q]\!]_{Rep(D)}$.*

We are now ready to prove our main result.

THEOREM 6.14. *The triples $\langle \mathcal{D}, Rep, \mathcal{Q}_{AUF}^{C'} \rangle$ and $\langle \mathcal{D}, Rep, \mathcal{Q}_{WD}^{C'} \rangle$ are representation systems.*

Since SELECT queries in SPARQL take as input an RDF graph but return a set of mappings (i.e., we do not have closure), it is not clear how to include them in the developed concept of a representation system (but see the discussion about SELECT in Section 7 below).

## 7. CERTAIN ANSWER COMPUTATION

This section studies how the certain answer to a SPARQL query $q$ over an RDF$^i$ database $D$ can be computed, i.e., how to compute $\bigcap [\![q]\!]_{Rep(D)}$. Having Theorem 6.14, it is easy to compute the certain answer to a query in the fragment of SPARQL $\mathcal{Q}_{AUF}^{C'}$ or $\mathcal{Q}_{WD}^{C'}$. Since $\langle \mathcal{D}, Rep, \mathcal{Q}_{AUF}^{C'} \rangle$ and $\langle \mathcal{D}, Rep, \mathcal{Q}_{WD}^{C'} \rangle$ are representation systems, we can apply Definition 6.2 for the identity query to get $\bigcap [\![q]\!]_{Rep(D)} = \bigcap Rep([\![q]\!]_D)$ for all $q$ and $D$. Thus, we can equivalently compute $\bigcap Rep([\![q]\!]_D)$ where $[\![q]\!]_D$ can be computed using the algebra of Section 5.

Before presenting the algorithm for certain answer computation, we need to introduce some auxiliary constructs similar to the ones defined in [12, 6] in the case of incomplete relational databases.

DEFINITION 7.1. *Let $D = (G, \phi)$ be an* RDF$^i$ *database. The* EQ-completed *form of $D$ is the* RDF$^i$ *database $D^{EQ} = (G^{EQ}, \phi)$ where $G^{EQ}$ is the same as $G$ except that all e-literals $\_l \in U$ appearing in $G$ have been replaced in $G^{EQ}$ by the constant $c \in C$ such that $\phi \models \_l$ EQ $c$ (if such a constant exists).*

In other words, in the EQ-completed form of an RDF$^i$ database $D$, all e-literals that are entailed by the global constraint to be equal to a constant from $C$ are substituted by that constant in all the triples in which they appear.

DEFINITION 7.2. *Let $D = (G, \phi)$ be an* RDF$^i$ *database. The* normalized *form of $D$ is the* RDF$^i$ *database $D^* = (G^*, \phi)$ where $G^*$ is the set*

$$\{(t, \theta) \mid (t, \theta_i) \in G \text{ for all } i = 1 \ldots n, \text{ and } \theta \text{ is } \bigvee_i \theta_i\}.$$

Given the above definition, the normalized form of an RDF$^i$ database $D$ is one that consists of the same global constraint and a graph in which conditional triples with the same triple part have been joined into a *single conditional triple* with a condition which is the disjunction of the conditions of the original triples. Notice that these new conditional triples do not follow Definition 3.1 which assumes conditions to be conjunctions of $\mathcal{L}$-constraints. We will allow this deviation from Definition 3.1 in this section.

LEMMA 7.3. *Let $D = (G, \phi)$ be an* RDF$^i$ *database. Then:*

$$\bigcap Rep(D) = \bigcap Rep((D^{EQ})^*)$$

Having Lemma 7.3, it is easy to give an algorithm that computes the certain answer to a query.

THEOREM 7.4. *Let $D = (G, \phi)$ be an* RDF$^i$ *database and $q$ a query from $\mathcal{Q}_{AUF}^{C'}$ or $\mathcal{Q}_{WD}^{C'}$. The certain answer of $q$ over $D$ can be computed as follows: i) compute $[\![q]\!]_D$ according to Section 5 and let $D_q = (G_q, \phi)$ be the resulting* RDF$^i$ *database, ii) compute the* RDF$^i$ *database $(H_q, \phi) = ((D_q)^{EQ})^*$, and iii) return the following set of RDF triples:*

$$\{(s, p, o) \mid ((s, p, o), \theta) \in H_q \text{ such that } \phi \models \theta \text{ and } o \notin U\}$$

Let us now present a preliminary analysis of the data complexity of computing the certain answer to a CONSTRUCT query over an RDF$^i$ database when $\mathcal{L}$ is a constraint language. Following [6], we first define the corresponding decision problem.

DEFINITION 7.5. *Let $q$ be a* CONSTRUCT *query. The* certainty problem *for query $q$, RDF graph $H$, and* RDF$^i$ *database $D$, is to decide whether $H \subseteq \bigcap [\![q]\!]_{Rep(D)}$. We denote this problem by $CERT_C(q, H, D)$.*

The next theorem shows how one can transform the certainty problem we defined above to the problem of deciding whether $\psi \in Th(\mathbf{M}_{\mathcal{L}})$ for an appropriate sentence $\psi$ of $\mathcal{L}$.

THEOREM 7.6. *Let $D = (G, \phi)$ be an* RDF$^i$ *database, $q$ a query from $\mathcal{Q}_{AUF}^{C'}$ or $\mathcal{Q}_{WD}^{C'}$, and $H$ an RDF graph. Then, $CERT_C(q, H, D)$ is equivalent to deciding whether the following formula is true in $\mathbf{M}_{\mathcal{L}}$:*

$$\bigwedge_{t \in H} (\forall \_l)(\phi(\_l) \supset \Theta(t, q, D, \_l)) \qquad (2)$$

*In the above formula:*

- *$\_l$ is the vector of all e-literals in the database $D$.*

- *$\Theta(t, q, D, \_l)$ is a disjunction $\theta_1 \vee \cdots \vee \theta_k$ that is constructed as follows. Let $[\![q]\!]_D = (G', \phi)$. $\Theta(t, q, D, \_l)$ has a disjunct $\theta_i$ for each conditional triple $(t_i', \theta_i') \in G'$ such that $t$ and $t_i'$ have the same subject and predicate. $\theta_i$ is:*

    - *$\theta_i'$ if $t$ and $t_i'$ have the same object as well.*
    - *$\theta_i' \wedge (\_l$ EQ $o)$ if the object of $t$ is $o \in C$ and the object of $t_i'$ is $\_l \in U$.*

    *If $t$ does not agree in the subject and predicate position with some $t_i'$, then $\Theta(t, q, D, \_l)$ is taken to be false.*

We can also prove a theorem like the above for SELECT queries by defining the relevant decision problem and developing appropriate versions of the relevant results of Section 6 that are needed. This involves first modifying Definition 6.2 so that $\mathcal{H}$ and $\mathcal{G}$ are sets of sets of mappings and $q$ is a SELECT query form (we call this SELECT-equivalence). Then, the condition of Definition 6.3, modified so that $\mathcal{Q}$-equivalence is substituted by SELECT-equivalence, can be proved using essentially the same techniques as the ones used to prove Theorem C.1.

## 7.1 Data Complexity Results

The data complexity of the certainty problem, $CERT_C(q, H, D)$, for $q$ in the $\mathcal{Q}_{AUF}^{C'}$ fragment of SPARQL and $D$ in the set of RDF$^i$ databases with constraints from ECL, diPCL, dePCL, and RCL is coNP-complete. This follows easily from known results of [6] for ECL and [14, 16, 35] for diPCL, dePCL, and RCL. Thus, we have the expected increase in data complexity given that the complexity of evaluating AND, UNION, and FILTER graph patterns over RDF graphs can be done in LOGSPACE [26].

Theorem 7.6 gives us immediately some easy upper bounds on the data complexity of the certainty problem in the case of RDF$^i$ with $\mathcal{L}$ equal to TCL or PCL. The satisfiability problem for conjunctions of TCL-constraints is known to be in PTIME [29]. Thus, the entailment problems arising in Theorem 7.6 can be trivially solved in EXPTIME. Therefore, the certainty problem is also in EXPTIME. To the best of our knowledge, no better bounds are known in the literature of TCL that we could have used to achieve a tighter bound for the certainty problem as we have done with the languages of the previous paragraph.

[20] shows that conjunctions of atomic RCC-5 constraints involving constants that are polygons in $V$-representation (called landmarks in [20]) can be decided in PTIME. Therefore, by restricting PCL so that only RCC-5 constraints are allowed and constants are given in $V$-representation, the certainty problem in this case is also in EXPTIME.

## 8. RELATED WORK

Incomplete information has been studied in-depth in relational databases starting with the paper of [12]. More recently, papers on uncertain [30, 3] and probabilistic [33] database models have reignited interest in this area.

In the context of the Web, incomplete information has been studied in detail for XML [2, 5]. Related work for incomplete information in RDF [9, 11, 4] has been discussed in the introduction, so we do not repeat the details here. The study of incomplete information in RDF undertaken in this paper goes beyond [4] where only the issue of OWA for RDF is investigated. Other cases of incomplete information in RDF (e.g., blank nodes according to the W3C RDF semantics which is different than the SPARQL semantics as we pointed out in Section 5) can also be investigated using an approach similar to ours. Comparing our work with [9, 11], we point out that these papers study complementary issues in the sense that they concentrate on temporal information of a specific kind only (validity time for a tuple). From a technical point of view, the approach of [11] is similar to ours since it is based on constraints, but, whereas we concentrate on query answering for $RDF^i$, [11] concentrates more on semantic issues such as temporal graph entailment. It is easy to see that $RDF^i$ can be used to represent incomplete temporal information that can be modeled as the object of a triple using any of the temporal constraint languages of [15]. An example of this situation is when we want to represent incomplete information about the time an event occurred. This is called *user-defined* time in the temporal database literature and it has not been studied in [9, 11].

Recently, some papers have started studying the problem of representing probabilistic information in RDF and querying it using SPARQL [34, 19]. It would be interesting to investigate how these approaches can be combined with the work presented in this paper as [8] has done in the model of probabilistic c-tables.

It is interesting to compare the expressive power that $RDF^i$ gives us to other recent works that use Semantic Web data models and languages for geospatial applications. When equipped with a constraint language like TCL, PCL, or RCL, $RDF^i$ goes beyond the proposals of the geospatial extensions of SPARQL, stSPARQL [18] and GeoSPARQL [23] that cannot query incomplete geospatial information. While GeoSPARQL provides a vocabulary for asserting topological relations (the topology vocabulary extension), the complexity of query evaluation over RDF graphs in this case has not been investigated so far in any detail and remains an open problem.

Incomplete geospatial information as it is studied in this paper can also be expressed in spatial description logics [24, 21]. For efficiency reasons, spatial DL reasoners such as RacerPro[2] and PelletSpatial [32] have opted for separating spatial relations from standard DL axioms as we have done by separating graphs and constraints. Since RDF graphs can be seen as DL ABoxes with atomic concepts only, all the results of this paper can be transferred to the relevant subsets of spatial DLs and their reasoners so they are of interest to this important Semantic Web area as well.

## 9. FUTURE WORK

Our future work focuses on the following: 1) explore other fragments of SPARQL that can be used to define a representation system for $RDF^i$, 2) study in more depth the complexity of certain answer computation for the various spatial and temporal constraint languages $\mathcal{L}$ we considered or the one used in [11] and identify tractable classes, and 3) study the complexity of evaluating various fragments of SPARQL over $RDF^i$ databases like it has been done in [26, 31] for the case of SPARQL and RDF.

---

[2] http://www.racer-systems.com/

## 10. REFERENCES

[1] S. Abiteboul, P. Kanellakis, and G. Grahne. On the Representation and Querying of Sets of Possible Worlds. In *SIGMOD*, pages 34–48, 1987.

[2] S. Abiteboul, L. Segoufin, and V. Vianu. Representing and querying XML with incomplete information. *ACM TODS*, 31(1):208–254, 2006.

[3] L. Antova, C. Koch, and D. Olteanu. $10^{(10^6)}$ worlds and beyond: efficient representation and processing of incomplete information. *VLDBJ*, 18(5):1021–1040, 2009.

[4] M. Arenas and J. Pérez. Querying semantic web data with SPARQL. In *PODS*, pages 305–316, 2011.

[5] P. Barceló, L. Libkin, A. Poggi, and C. Sirangelo. XML with incomplete information. *JACM*, 58(1):4, 2010.

[6] G. Grahne. *The Problem of Incomplete Information in Relational Databases*. LNCS. Springer Verlag, 1991.

[7] G. Grahne. Incomplete information. In *Encyclopedia of Database Systems*, pages 1405–1410. Springer, 2009.

[8] T. J. Green and V. Tannen. Models for incomplete and probabilistic information. In *EDBT Workshops*, 2006.

[9] C. Gutierrez, C. A. Hurtado, and A. A. Vaisman. Introducing Time into RDF. *IEEE TKDE*, 19(2), 2007.

[10] P. Hayes. RDF Semantics, W3C Recommendation, 2004.

[11] C. A. Hurtado and A. A. Vaisman. Reasoning with Temporal Constraints in RDF. In *PPSWR*. Springer, 2006.

[12] T. Imielinski and W. Lipski. Incomplete Information in Relational Databases. *JACM*, 31(4):761–791, 1984.

[13] P. C. Kanellakis, G. M. Kuper, and P. Z. Revesz. Constraint Query Languages. In *PODS*, 1990.

[14] M. Koubarakis. Complexity results for first-order theories of temporal constraints. In *KR*, pages 379–390, 1994.

[15] M. Koubarakis. Database models for infinite and indefinite temporal information. *Inf. Syst.*, 19(2):141–173, 1994.

[16] M. Koubarakis. The complexity of query evaluation in indefinite temporal constraint databases. *Theor. Comput. Sci.*, 171(1-2):25–60, 1997.

[17] M. Koubarakis and K. Kyzirakos. Modeling and querying metadata in the semantic sensor web: The model stRDF and the query language stSPARQL. In *ESWC*, 2010.

[18] K. Kyzirakos, M. Karpathiotakis, and M. Koubarakis. Strabon: A Semantic Geospatial DBMS. In *ISWC'12*, pages 295–311, 2012.

[19] X. Lian and L. Chen. Efficient query answering in probabilistic rdf graphs. In *SIGMOD*, pages 157–168,

2011.

[20] W. Liu, S. Wang, S. Li, and D. Liu. Solving qualitative constraints involving landmarks. In *CP*, 2011.

[21] C. Lutz and M. Miličić. A tableau algorithm for description logics with concrete domains and general tboxes. *J. Autom. Reason.*, 38:227–259, April 2007.

[22] C. Nikolaou and M. Koubarakis. Querying Linked Geospatial Data with Incomplete Information. In *5th International Terra Cognita Workshop - Foundations, Technologies and Applications of the Geospatial Web*, Boston, USA, 2012.

[23] Open Geospatial Consortium Inc. GeoSPARQL - A geographic query language for RDF data. OGC, 2010.

[24] Ö. Özcep and R. Möller. Computationally feasible query answering over spatio-thematic ontologies. In *GEOProcessing*, 2012.

[25] J. Pérez, M. Arenas, and C. Gutierrez. Semantics of SPARQL. Technical report, Univ. de Chile, 2006.

[26] J. Pérez, M. Arenas, and C. Gutierrez. Semantics and complexity of SPARQL. *ACM TODS*, 34(3):1–45, 2009.

[27] E. Prud'hommeaux and A. Seaborne. SPARQL Query Language for RDF. W3C Recommendation 15 Jan. 2008.

[28] D. A. Randell, Z. Cui, and A. G. Cohn. A spatial logic based on regions and connection. In *KR*, 1992.

[29] J. Renz and B. Nebel. On the complexity of qualitative spatial reasoning: A maximal tractable fragment of the region connection calculus. *AIJ*, 108(1-2):69–123, 1999.

[30] A. D. Sarma, O. Benjelloun, A. Y. Halevy, and J. Widom. Working models for uncertain data. In *ICDE*, 2006.

[31] M. Schmidt, M. Meier, and G. Lausen. Foundations of sparql query optimization. In *ICDT*, pages 4–33, 2010.

[32] M. Stocker and E. Sirin. PelletSpatial: A Hybrid RCC-8 and RDF/OWL Reasoning and Query Engine. In *OWLED*, 2009.

[33] D. Suciu, D. Olteanu, C. Ré, and C. Koch. *Probabilistic Databases*. Morgan & Claypool Publishers, 2011.

[34] O. Udrea, V. S. Subrahmanian, and Z. Majkic. Probabilistic RDF. In *IRI*, pages 172–177, 2006.

[35] R. van der Meyden. The complexity of querying indefinite data about linearly ordered domains. In *PODS*, pages 331–345, 1992.

# APPENDIX

The Appendix is structured as follows. In Section A we formally define a number of constraint languages for modeling information in geospatial and temporal domains. These languages have been already defined informally in Section 2. Then, Section B gives formal definitions for the concept of *well-designed* graph patterns and relevant concepts, such as *subsumption for mappings* and *weak monotonicity*, while it presents known results for well-designed patterns. These results are useful for establishing the monotonicity results for the fragment of SPARQL corresponding to CONSTRUCT queries with well-designed graph patterns and without blank nodes in their templates. Section C provides the proofs for the section of Representation Systems (Section 6), while Section D provides the proofs for the section of Certain Answer Computation (Section 7). Last, Section E is devoted to additional propositions that are useful to establish some propositions and/or theorems of Section 6.

## A. CONSTRAINT LANGUAGES

In this section we define formally all the constraint languages used in the paper.

### A.1 The Language ECL

The language *ECL* (*E*quality *C*onstraint *L*anguage) with predicate symbol $=$ and an infinite number of constants has been defined in [13]. The intended structure for this language interprets symbol $=$ as equality and constants as "themselves". An ECL-constraint is an ECL formula of the form $x_1 = x_2$ where $x_1, x_2$ are variables or constants.

ECL has been used by [13] for the development of an extended relational model based on ECL-constraints and by [12, 1, 6] for querying and updating incomplete information in relational databases.

The following two languages are from [16].

### A.2 The Language diPCL

The language *diPCL* (*di*screte *P*oint *C*onstraint *L*anguage) allows us to make statements about points in discrete time. diPCL is a first-order language with constants from the set of integers $\mathbb{Z}$, a 2-ary function symbol $-$, and a binary predicate symbol $<$. The terms and atomic formulae of diPCL are defined as follows. Constants and variables are *terms*. If $t_1$ and $t_2$ are constants or variables, then $t_1 - t_2$ is a term. An *atomic formula* of diPCL (diPCL-constraint) is a formula of the form $t \sim c$ or $c \sim t$ where $\sim$ is either $<$ or $=$, $t$ is a term, and $c$ is a constant. For example, the following are diPCL-constraints:

$$x_1 - x_2 < 2,\ x_1 = 5,\ x_1 < 6$$

The intended structure for diPCL, denoted by $\mathbf{M}_{diPCL}$, has the set of integer constants as its domain. $\mathbf{M}_{diPCL}$ interprets each constant symbol by the corresponding integer number in $\mathbb{N}$, function symbol $-$ by the subtraction operation over the integers, and predicate symbol $<$ by the relation "less than". Then, theory $Th(\mathbf{M}_{diPCL})$ is a sub-theory of $Th(\mathbb{Z}, +, <)$, the theory of integers with addition and order.

### A.3 The Language dePCL

The language *dePCL* (*de*nse *P*oint *C*onstraint *L*anguage) allows us to make statements about points in dense time.

dePCL is a first-order language with constants from the set of rational numbers $\mathbb{Q}$, a 2-ary function symbol $-$, and a binary predicate symbol $<$. The terms and atomic formulae of dePCL are defined as follows. Constants and variables are *terms*. If $t_1$ and $t_2$ are constants or variables, then $t_1 - t_2$ is a term. An *atomic formula* of dePCL (dePCL-constraint) is a formula of the form $t \sim c$ or $c \sim t$ where $\sim$ is either $<$ or $=$, $t$ is a term, and $c$ is a constant. For example, the following are dePCL-constraints:

$$x_1 - x_2 < 1/2, \ x_1 = 5/2, \ x_1 < 6/1$$

The intended structure for dePCL, denoted by $\mathbf{M}_{dePCL}$, has the set of rational constants as its domain. $\mathbf{M}_{dePCL}$ interprets each constant symbol by the corresponding rational number in $\mathbb{Q}$, function symbol $-$ by the subtraction operation over the rationals, and predicate sybmol $<$ by the relation "less than". Then, theory $Th(\mathbf{M}_{dePCL})$ is a sub-theory of $Th(\mathbb{R}, +, <)$, the theory of real numbers with addition and order.

### A.4 The Language TCL

The language *TCL* (*T*opological *C*onstraint *L*anguage) allows us to represent topological properties of non-empty, regular closed subsets of $\mathbb{Q}^2$ (we will call these subsets *regions* for brevity). TCL is a first-order language with the following 6 binary predicate symbols: DC, EC, PO, EQ, TPP, and NTPP. An *atomic formula* of TCL is a formula of the form $r_1 \ R \ r_2$ where $r_1, r_2$ are variables and $R$ is one of the above predicates. We will often use the terminology $\mathcal{L}$-*constraints* to refer to atomic formulae of an arbitrary constraint language $\mathcal{L}$. For example, the following are TCL-constraints:

$$r_1 \text{ NTPP } r_2, \ r_2 \text{ PO } r_3, \ r_2 \text{ EQ } r_3$$

The intended structure for TCL, denoted by $\mathbf{M}_{TCL}$, has the set of regions as its domain, and interprets each of the predicate symbols given above by the corresponding topological relation of RCC-8 [28]. Note that relations NTPPi and TPPi of RCC-8 are not included in the vocabulary of TCL since they can be expressed by interchanging the arguments of NTPP and TPP.

The language TCL allows us to capture the topology of regions of interest to an application but makes no commitment regarding other non-topological properties of these regions, e.g., shape. The language PCL considered below deals with polygonal shapes.

### A.5 The Language RCL

The language *RCL* (*R*ectangle *C*onstraint *L*anguage) allows us to capture spatial and metric constraints (e.g., topological or directional, and horizontal or vertical distance constraints among the edges of rectangles) involving rectangles with sides parallel to the axes in $\mathbb{Q}^2$ (we will call them *boxes*). RCL is useful not only for modeling regions of space with such rectangular shapes but also for modeling *minimum bounding rectangles* that are typically used as approximations of spatial objects, e.g., in spatial data structures and elsewhere.

RCL is a first-order language with equality and 2 sorts: the sort $\mathcal{Q}$ for rational constants, and the sort $\mathcal{R}$ for boxes. The set of non-logical symbols of RCL includes: all rational constants of sort $\mathcal{Q}$, a 2-ary function symbol $-$ of sort $(\mathcal{Q}, \mathcal{Q}, \mathcal{Q})$, function symbols $LL_x(\cdot), LL_y(\cdot), UR_x(\cdot), UR_y(\cdot)$ of sort $(\mathcal{R}, \mathcal{Q})$, and predicate symbol $<$ of sort $(\mathcal{Q}, \mathcal{Q})$.

The terms and atomic formulae of RCL are defined as follows. Constants of sort $\mathcal{Q}$ and variables of sort $\mathcal{R}$ are terms. If $r$ is a variable of sort $\mathcal{R}$ then $LL_x(r), LL_y(r), UR_x(r)$ and $UR_y(r)$ is a term of sort $\mathcal{Q}$. If $t_1, t_2$ are terms of sort $\mathcal{Q}$, then $t_1 - t_2$ is a term of sort $\mathcal{Q}$. An *atomic formula* of RCL is a formula of the form $t \sim c$ where $\sim$ is $<$ or $=$, $t$ is a term of sort $\mathcal{Q}$, and $c$ a rational constant. Symbol $=$ is the equality predicate for sort $\mathcal{Q}$; we will not use the equality predicate for sort $\mathcal{R}$ in our formulae.

The intended structure for RCL, denoted by $\mathbf{M}_{RCL}$, interprets each non-logical symbol as follows. Each rational constant is interpreted by its corresponding rational number. The function symbol $-$ is interpreted by the subtraction operation over the rationals, while the function symbols $LL_x(\cdot), LL_y(\cdot), UR_x(\cdot)$ and $UR_y(\cdot)$ are interpreted by the easily-defined functions that given a box in $\mathbb{Q}^2$, return the $x$- and $y$-coordinate of its lower-left and lower-right vertex respectively. Predicate $<$ is interpreted by the relation "less than" over $\mathbb{Q}$.

A *RCL-constraint* is a RCL formula of the form $t \sim c$ where $\sim$ is $=, <, >, \leq$ or $\geq$, $t$ is a term of sort $\mathcal{Q}$, and $c$ a rational constant (the predicates $<, \leq,$ and $\geq$ are defined as usual). For example, the following are RCL-constraints:

$$LL_x(r_2) - LL_x(r_1) < 0, \ UR_y(r_1) - LL_y(r_2) = 5/2$$

## B. WELL-DESIGNED GRAPH PATTERNS

In this section we present relevant material and known results for the fragment of SPARQL corresponding to the notion of well-designed graph patterns. These come from [26, 4].

The next definition introduces the notion of well-designed graph patterns.

DEFINITION B.1 (WELL-DESIGNED PATTERNS [4]).
*Let $P$ be a graph pattern in the AND-FILTER-OPT fragment of SPARQL. Then $P$ is* well-designed *if (1) $P$ is* safe, *i.e., for every sub-pattern $(P_1 \ FILTER \ R)$ of $P$, it holds that $var(R) \subseteq var(P_1)$, and (2) for every sub-pattern $P' = (P_1 \ OPT \ P_2)$ of $P$ and variable $?X$, if $?X$ occurs both inside $P_2$ and outside $P'$, then it also occurs in $P_1$.*

In [26, 4], the authors identified in well-designed graph patterns unique and interesting properties that make query evaluation more efficient in contrast to what you get without the syntactic restrictions imposed on the graph patterns by Definition B.1 above. One of these properties is that the fragment of SPARQL graph patterns corresponding to well-designed graph patterns is *weakly monotone*. In the following we introduce the notion of *weak monotonicity*, but first we define the notion of *subsumption* for mappings which is needed for *weak monotonicity*.

DEFINITION B.2 (SUBSUMPTION OF MAPPINGS). *Let $\mu_1, \mu_2$ be mappings. We say that $\mu_1$ is subsumed by $\mu_2$, denoted by $\mu_1 \preceq \mu_2$, if $dom(\mu_1) \subseteq dom(\mu_2)$ and $\mu_1(x) = \mu_2(x)$ for every $x \in dom(\mu_1)$. Let $\Omega_1, \Omega_2$ be set of mappings. We say that $\Omega_1$ is subsumed by $\Omega_2$, denoted by $\Omega_1 \sqsubseteq \Omega_2$, if for every $\mu_1 \in \Omega_1$ there exists mapping $\mu_2 \in \Omega_2$ such that $\mu_1 \preceq \mu_2$.*

EXAMPLE B.3. *Let us consider Example 5.2 again. Mapping $\mu_4$ is subsumed by mapping $\mu_1$, i.e., $\mu_4 \preceq \mu_1$.*

*Informally, when a mapping $\mu$ subsumes a mapping $\mu'$, then $\mu$ contains additional information to $\mu'$, i.e., it maps additional variables to RDF terms.*

DEFINITION B.4 (WEAK MONOTONICITY). *Let $P$ be a graph pattern of SPARQL. $P$ is said to be* weakly monotone *if for every pair $G, H$ of RDF graphs such that $G \subseteq H$, it is $[\![P]\!]_G \sqsubseteq [\![P]\!]_H$.*

From [4] we know that every well-designed graph pattern is weakly monotone.

THEOREM B.5 (THEOREM 4.3 OF [4]). *Every well-designed graph pattern is weakly monotone.*

## C. PROOFS FOR SECTION 6

### C.1 Proof of Proposition 6.5

Proof for part *a*)
The monotonicity property for $\mathcal{Q}_{AUF}^S$ follows easily from the monotonicity property of graph patterns containing only the AND, UNION, and FILTER operators as presented in [4, Lemma 3.2].

Proof for part *b*)
From the same paper, it trivially follows that $\mathcal{Q}_{OPT}^S$ and $\mathcal{Q}_{OPT}^C$ are not monotone.

Proof for part *c*)
Now consider a query $q = (E, P) \in \mathcal{Q}_{AUF}^C$ and let $G, H$ be two RDF graphs such that $G \subseteq H$. According to Definition 4.6 of CONSTRUCT for RDF graphs as given in [25] we have

$$[\![q]\!]_G = \bigcup_{\mu \in [\![P]\!]_G} \{\mu(f_\mu(E)) \cap ((I \cup B) \times I \times T)\} \quad (1)$$

$$[\![q]\!]_H = \bigcup_{\mu' \in [\![P]\!]_H} \{\mu'(f_{\mu'}(E)) \cap ((I \cup B) \times I \times T)\} \quad (2)$$

From the monotonicity property of AUF graph patterns [4] we have that $[\![P]\!]_G \subseteq [\![P]\!]_H$. Therefore, all mappings $\mu$ appearing in the union expression of formula (1) appear also in the union expression of formula (2). Therefore, if the sets in formulae (1) and (2) are the same, then we shall get the required relation for monotonicity, that is, $[\![q]\!]_G \subseteq [\![q]\!]_H$.

Notice, however, that this is not the case because of the renaming functions. According to Definition 4.5 of [25], a renaming function not only depends on the mapping that has been constructed from the evaluation of a graph pattern, but also on the underlying RDF graph over which the graph pattern is evaluated. Thus, a renaming function besides renaming a specific blank node to another one per each mapping solution, that renaming has to correspond to a fresh blank node not appearing in the underlying RDF graph. Therefore, a renaming function used in formula (1) could have possibly renamed a blank node to a fresh one regarding $G$, but not a fresh one regarding $H$, i.e., that blank node could have been already in $H$.

Hence, in order to have the monotonicity property for $\mathcal{Q}_{AUF}^C$, we have to restrict ourselves in CONSTRUCT queries without blank nodes in their template. In such a case, the renaming functions do not have any effect on the templates of CONSTRUCT queries. Hence, the sets in formulae (1) and (2) are the same for same mappings, and thus,

$$[\![q]\!]_G \subseteq [\![q]\!]_H.$$

Proof for part *d*)
Consider a query $q = (E, P) \in \mathcal{Q}_{WD}^{C'}$ and let $G, H$ be two RDF graphs such that $G \subseteq H$. According to Definition 4.6 of CONSTRUCT for RDF graphs as given in [25] we have

$$[\![q]\!]_G = \bigcup_{\mu \in [\![P]\!]_G} \{\mu(f_\mu(E)) \cap ((I \cup B) \times I \times T)\} \quad (3)$$

$$[\![q]\!]_H = \bigcup_{\mu' \in [\![P]\!]_H} \{\mu'(f_{\mu'}(E)) \cap ((I \cup B) \times I \times T)\} \quad (4)$$

Since the template $E$ does not contain blank nodes, we can omit the renaming functions from these expressions and get

$$[\![q]\!]_G = \bigcup_{\mu \in [\![P]\!]_G} \{\mu(E) \cap ((I \cup B) \times I \times T)\} \quad (5)$$

$$[\![q]\!]_H = \bigcup_{\mu' \in [\![P]\!]_H} \{\mu'(E) \cap ((I \cup B) \times I \times T)\} \quad (6)$$

Since $P$ is well-designed, it follows from Theorem B.5 that $P$ is weakly monotone. Therefore, $[\![P]\!]_G \sqsubseteq [\![P]\!]_H$. Hence, for every mapping $\mu \in [\![P]\!]_G$ there exists mapping $\mu' \in [\![P]\!]_H$ such that $\mu \preceq \mu'$. This means that $\mu, \mu'$ map the common variables of their domains to the same RDF terms. Hence, if a mapping $\mu \in [\![P]\!]_G$ produces triple $t$ in $[\![q]\!]_G$, that triple is also produced in $[\![q]\!]_H$ by a mapping $\mu' \in [\![P]\!]_H$ such that $\mu \preceq \mu'$. Thus, $[\![q]\!]_G \subseteq [\![q]\!]_H$.

### C.2 Proof of Proposition 6.8

The proof of Proposition 6.8 is straightforward from the monotonicity property. Since $\mathcal{G} \approx \mathcal{H}$ we have the following:

- for every $G \in \mathcal{G}$ there exists $H \in \mathcal{H}$ such that $H \subseteq G$ and
- for every $H \in \mathcal{H}$ there exists $G \in \mathcal{G}$ such that $G \subseteq H$.

Let $q \in \mathcal{Q}$. Because $\mathcal{Q}$ is monotone, from the first item above we have that $[\![q]\!]_H \subseteq [\![q]\!]_G$ for every $G \in \mathcal{G}$ and some $H \in \mathcal{H}$. Notice also that this property holds for every set making up $[\![q]\!]_\mathcal{G}$ and some $[\![q]\!]_H \in [\![q]\!]_\mathcal{H}$. Similarly, from the second item above we have that $[\![q]\!]_G \subseteq [\![q]\!]_H$ for every $H \in \mathcal{H}$ and some $G \in \mathcal{G}$. Notice again that this property holds for every set making up $[\![q]\!]_\mathcal{H}$ and some $[\![q]\!]_G \in [\![q]\!]_\mathcal{G}$. Hence, $[\![q]\!]_\mathcal{G}$ and $[\![q]\!]_\mathcal{H}$ are coinitial, that is,

$$[\![q]\!]_\mathcal{G} \approx [\![q]\!]_\mathcal{H}.$$

### C.3 Proof of Lemma 6.9

The proof is similar to the one given in [12, Lemma 4.2]. We have to prove that $\bigcap [\![q]\!]_\mathcal{G} = \bigcap [\![q]\!]_\mathcal{H}$ for every $q \in \mathcal{Q}_{AUF}^{C'}$. Let $\mathcal{G} \approx \mathcal{H}$. Then, from Proposition 6.8 and because of the monotonicity property of $\mathcal{Q}_{AUF}^{C'}$, we have that $[\![q]\!]_\mathcal{G} \approx [\![q]\!]_\mathcal{H}$ for every $q \in \mathcal{Q}_{AUF}^{C'}$. Thus, for every $[\![q]\!]_G \in [\![q]\!]_\mathcal{G}$ there exists an $[\![q]\!]_{H_G} \in [\![q]\!]_\mathcal{H}$ such that $[\![q]\!]_{H_G} \subseteq [\![q]\!]_G$. So, we have

$$\bigcap [\![q]\!]_\mathcal{G} = \bigcap_{G \in \mathcal{G}} [\![q]\!]_G \supseteq \bigcap_{G \in \mathcal{G}} [\![q]\!]_{H_G} \supseteq \bigcap_{H \in \mathcal{H}} [\![q]\!]_H = \bigcap [\![q]\!]_\mathcal{H}.$$

To see why $\bigcap_{G \in \mathcal{G}} [\![q]\!]_G \supseteq \bigcap_{G \in \mathcal{G}} [\![q]\!]_{H_G}$, notice that $\bigcap_{G \in \mathcal{G}} [\![q]\!]_G$ and $\bigcap_{G \in \mathcal{G}} [\![q]\!]_{H_G}$ can be written respectively as

$$[\![q]\!]_{G_1} \cap [\![q]\!]_{G_2} \cap \cdots \quad \text{and} \quad [\![q]\!]_{H_{G_1}} \cap [\![q]\!]_{H_{G_2}} \cap \cdots$$

and that
$$[\![q]\!]_{H_{G_i}} \subseteq [\![q]\!]_{G_i}.$$

Therefore, if an element $x$ is in $\bigcap_{G \in \mathcal{G}} [\![q]\!]_{H_G}$, it will be in every $[\![q]\!]_{H_{G_i}}$, and thus it will be in every $[\![q]\!]_{G_i}$, which proves the relation.

Now, to see why $\bigcap_{G \in \mathcal{G}} [\![q]\!]_{H_G} \supseteq \bigcap_{H \in \mathcal{H}} [\![q]\!]_H$, notice that the relation can be written as
$$\bigcap [\![q]\!]_{\mathcal{H}_{\mathcal{G}}} \supseteq \bigcap [\![q]\!]_{\mathcal{H}}$$
where
$$\mathcal{H}_{\mathcal{G}} \equiv \{H \in \mathcal{H} \mid H \subseteq G \text{ for some } G \in \mathcal{G}\}.$$

Thus, $\mathcal{H}_{\mathcal{G}} \subseteq \mathcal{H}$, and therefore, we have that $\bigcap \mathcal{H}_{\mathcal{G}} \supseteq \bigcap \mathcal{H}$. Similarly if $q$ is a monotone query, we have $[\![q]\!]_{\mathcal{H}_{\mathcal{G}}} \subseteq [\![q]\!]_{\mathcal{H}}$ and $\bigcap [\![q]\!]_{\mathcal{H}_{\mathcal{G}}} \supseteq \bigcap [\![q]\!]_{\mathcal{H}}$.

Therefore, we showed that
$$\bigcap [\![q]\!]_{\mathcal{G}} \supseteq \bigcap [\![q]\!]_{\mathcal{H}}.$$

We work similarly to prove
$$\bigcap [\![q]\!]_{\mathcal{H}} \supseteq \bigcap [\![q]\!]_{\mathcal{G}}$$
and get
$$\bigcap [\![q]\!]_{\mathcal{G}} = \bigcap [\![q]\!]_{\mathcal{H}}.$$

## C.4 Proof of Proposition 6.13

Let $[\![q]\!]_D$ be the pair $D_1 = (G_1, \phi)$ and $G'$ an RDF graph such that $G' \in Rep(D_1)$. From the definition of $Rep$, there exists a valuation $v'$ such that $\mathbf{M}_{\mathcal{L}} \models v'(\phi)$ and $G' \supseteq v'(G_1)$. From the assumption that $v([\![q]\!]_D) = [\![q]\!]_{v(D)}$ and since $\mathbf{M}_{\mathcal{L}} \models v'(\phi)$ we get
$$G' \supseteq v'(G_1) = v'(D_1) = v'([\![q]\!]_D) = [\![q]\!]_{v'(D)} = H$$
where $H$ is a new symbol introduced for convenience. Now, observe that $v'(D)$ is the RDF graph $v'(G)$ which is an element of $Rep(D)$ since $\mathbf{M}_{\mathcal{L}} \models v'(\phi)$. Since also
$$[\![q]\!]_{Rep(D)} = \{[\![q]\!]_G \mid G \in Rep(D)\}$$
it turns out that $H \in [\![q]\!]_{Rep(D)}$. To see this, notice that
$$H = [\![q]\!]_{v'(D)}$$
and that
$$v'(D) \in Rep(D).$$

This proves that for each $G' \in Rep([\![q]\!]_D)$ there exists an $H \in [\![q]\!]_{Rep(D)}$ such that $H \subseteq G'$. To prove that $Rep([\![q]\!]_D) \approx [\![q]\!]_{Rep(D)}$ we need to show the same for the other direction.

Let $H'$ be an RDF graph such that $H' \in [\![q]\!]_{Rep(D)}$. Then $H' = [\![q]\!]_H$ for some $H \in Rep(D)$. From the definition of $Rep$, there exists a valuation $v'$ such that $\mathbf{M}_{\mathcal{L}} \models v'(\phi)$ and $H \supseteq v'(G)$ or equivalently $H \supseteq v'(D)$. From our assumption that $v([\![q]\!]_D) = [\![q]\!]_{v(D)}$ and $\mathbf{M}_{\mathcal{L}} \models v'(\phi)$, we have $[\![q]\!]_{v'(D)} = v'([\![q]\!]_D)$. Since $q$ belongs to a monotone fragment of SPARQL and $H \supseteq v'(D)$, we have
$$[\![q]\!]_H \supseteq [\![q]\!]_{v'(D)}$$
which is equivalent to
$$H' \supseteq v'([\![q]\!]_D).$$

Now observe that since $\mathbf{M}_{\mathcal{L}} \models v'(\phi)$, $v'([\![q]\!]_D)$ is an RDF graph $G'$ and that $G' \in Rep([\![q]\!]_D)$. Therefore, we showed that for every $H' \in [\![q]\!]_{Rep(D)}$ there exists an $G' \in Rep([\![q]\!]_D)$ such that $G' \subseteq H'$.

Hence
$$Rep([\![q]\!]_D) \approx [\![q]\!]_{Rep(D)}.$$

## C.5 Proof of Theorem 6.14

The proof for item $a$) can be found in the proof for Theorem C.1, while the proof for item $b$) can be found in the proof for Theorem C.2 below.

THEOREM C.1. *The triple $\langle \mathcal{D}, Rep, \mathcal{Q}_{AUF}^{C'} \rangle$ is a representation system.*

PROOF. To prove Theorem C.1, it is sufficient to show that for any $D = (G, \phi) \in \mathcal{D}$ and any query $q = (E, P) \in \mathcal{Q}_{AUF}^{C'}$ it is possible to define $[\![q]\!]_D$ in such a way that
$$Rep([\![q]\!]_D) \equiv_{\mathcal{Q}_{AUF}^{C'}} [\![q]\!]_{Rep(D)}.$$

By Lemma 6.9 it is sufficient to prove that
$$Rep([\![q]\!]_D) \approx [\![q]\!]_{Rep(D)}. \tag{1}$$

From Proposition 6.13 it now suffices to prove that for any valuation $v$ such that $\mathbf{M}_{\mathcal{L}} \models v(\phi)$ it is
$$v([\![q]\!]_D) = [\![q]\!]_{v(D)}.$$

From Proposition E.2, the above holds if for any valuation $v$ such that $\mathbf{M}_{\mathcal{L}} \models v(\phi)$, the following holds
$$v([\![P]\!]_D) = [\![P]\!]_{v(D)}.$$

This is done by induction on the structure of graph patterns $P$ of $\mathcal{Q}_{AUF}^{C'}$.

- $P$ is $(s, p, o)$ (base case):

  We shall prove that $v([\![P]\!]_D) = [\![P]\!]_{v(D)}$.

  Let $\mu \in v([\![P]\!]_D)$. Then, there exists a conditional mapping $\mu' = (\nu', \theta') \in [\![P]\!]_D$ such that $v(\mu') = \mu$ and $\mathbf{M}_{\mathcal{L}} \models v(\theta')$.

  We now distinguish two cases corresponding to the two cases of Definition 5.15 (1):

  (i) In this case $o \in C$. In this case $dom(\nu')$ does not contain any special query variable, hence the application of $v$ to $\mu'$ leaves $\nu'$ unchanged. In other words $\mu = v(\mu') = \nu'$.

  Now we have two cases corresponding to the two sets making up $[\![P]\!]_D$.

  If $\mu = \nu'$ is an element of the first set, then
  $$(\mu'(P), \theta') \in G.$$
  Since $\mathbf{M}_{\mathcal{L}} \models v(\theta')$, this is written as
  $$v(\mu'(P)) \in v(G)$$
  and because $\mathbf{M}_{\mathcal{L}} \models v(\phi)$, this is equivalent to
  $$v(\mu'(P)) \in v(D).$$
  Since also $v(\mu') = \mu$, we have
  $$\mu(P) \in v(D).$$
  and hence
  $$\mu \in [\![P]\!]_{v(D)}.$$

If $\mu = \nu'$ is an element of the second set then $\theta'$ is $\theta \wedge (\_l \ EQ \ o)$. Since $\mathbf{M}_\mathcal{L} \models v(\theta')$, we have $\mathbf{M}_\mathcal{L} \models v(\theta)$ and $\mathbf{M}_\mathcal{L} \models v(\_l \ EQ \ o)$. From the second set of the first item of Definition 5.15 that makes up $[\![P]\!]_D$, we have

$$((\mu(s), \mu(p), \_l), \theta) \in G.$$

Since $\mathbf{M}_\mathcal{L} \models v(\theta)$, we can apply $v$ to the above and get

$$v((\mu(s), \mu(p), \_l)) \in v(G).$$

Since also $\mathbf{M}_\mathcal{L} \models v(\phi)$ and $\mathbf{M}_\mathcal{L} \models v(\_l \ EQ \ o)$ we get

$$(\mu(s), \mu(p), o) \in v(D)$$

which is equivalently written as

$$\mu(P) \in v(D)$$

or

$$\mu \in [\![P]\!]_{v(D)}.$$

(ii) In this case $o \in I \cup B \cup L \cup V$. Therefore $\nu'(o) \in I \cup B \cup L \cup U \cup C$ and

$$(\mu'(P), \theta') \in G.$$

Since $\mathbf{M}_\mathcal{L} \models v(\theta')$, we can apply $v$ to the previous relation and get

$$v(\mu'(P)) \in v(G).$$

Because also $\mathbf{M}_\mathcal{L} \models v(\phi)$, we have

$$v(\mu'(P)) \in v(D).$$

The latter fact together with the fact that $v(\mu') = \mu$ gives that

$$\mu(P) \in v(D)$$

and hence

$$\mu \in [\![P]\!]_{v(D)}.$$

This establishes the fact that $v([\![P]\!]_D) \subseteq [\![P]\!]_{v(D)}$. The other direction of the proof is similar and goes as follows.

Let $\mu \in [\![P]\!]_{v(D)}$. Then, $\mu(P) \in v(D)$.

We now distinguish two cases corresponding to the two cases of Definition 5.15 (1):

(i) In this case $o \in C$. Then, $dom(\mu)$ does not contain any special query variable. Since $\mu(P) \in v(D)$, there exists conditional triple $((\mu(s), \mu(p), x), \theta) \in G$ such that $\mathbf{M}_\mathcal{L} \models v(\theta)$ and $v(x) = o$.

Now, we have two cases for $x$ corresponding to the two sets making up $[\![P]\!]_D$ in Definition 5.15 (1):

- $x$ is $o$. Then, a conditional mapping $\mu' = (\mu, \theta)$ is an element of the first set, i.e., $\mu' \in [\![P]\!]_D$. Since $\mathbf{M}_\mathcal{L} \models v(\theta)$, we can apply $v$ to relation

$$\mu' \in [\![P]\!]_D$$

and get

$$v(\mu') \in v([\![P]\!]_D).$$

Because $dom(\mu)$ does not contain any special query variable the application of $v$ to $\mu'$ leaves $\mu'$ unchanged. Therefore,

$$v(\mu') \in v([\![P]\!]_D)$$

becomes

$$\mu \in v([\![P]\!]_D).$$

- $x$ is $\_l$. Then, a conditional mapping $\mu' = (\mu, \theta \wedge \_l \ EQ \ o)$ is an element of the second set, i.e., $\mu' \in [\![P]\!]_D$. Since $v(\_l) = v(x) = o$, we have $\mathbf{M}_\mathcal{L} \models v(\_l \ EQ \ o)$. Because also $\mathbf{M}_\mathcal{L} \models v(\theta)$, it holds that $\mathbf{M}_\mathcal{L} \models (\theta \wedge \_l \ EQ \ o)$, and hence we can apply $v$ to relation

$$\mu' \in [\![P]\!]_D$$

and get

$$v(\mu') \in v([\![P]\!]_D).$$

Because $dom(\mu)$ does not contain any special query variable the application of $v$ to $\mu'$ leaves $\mu'$ unchanged. Therefore,

$$v(\mu') \in v([\![P]\!]_D)$$

becomes

$$\mu \in v([\![P]\!]_D).$$

(ii) In this case $o \in I \cup B \cup L \cup V$. We have two cases to consider.

If $o \in I \cup B \cup L \cup V_n$, then $dom(\mu)$ does not contain any special query variable. Since $\mu(P) \in v(D)$, there exists conditional triple $(\mu(P), \theta) \in G$ such that $\mathbf{M}_\mathcal{L} \models v(\theta)$. By the "else" part of Definition 5.15 the conditional mapping $\mu' = (\mu, \theta)$ is an element of $[\![P]\!]_D$, that is,

$$\mu' \in [\![P]\!]_D.$$

Since $\mathbf{M}_\mathcal{L} \models v(\theta)$, we can apply valuation $v$ to this relation and get

$$v(\mu') \in v([\![P]\!]_D)$$

which is equivalent to

$$\mu \in v([\![P]\!]_D)$$

since the application of $v$ to $\mu'$ leaves $\mu'$ (and $\mu$) unchanged.

Now if $o \in V_s$, there exists a conditional mapping $\mu' = (\nu', \theta)$ such that $\mu'$ and $\mu$ are possibly compatible, $dom(\mu') = dom(\mu)$, and $\mathbf{M}_\mathcal{L} \models v(\theta)$. The conditional mapping $\mu'$ is such that either $\nu' = \mu$ or $\nu'(x) = \mu(x)$ for every $x \in dom(\mu) \setminus \{o\}$ and $\nu'(o) \in U$ with $v(\nu'(o)) = \mu(o)$. In either case $\mu(P) \in v(D)$ implies

$$v(\mu'(P)) \in v(D).$$

from which eliminating $v$ we get

$$(\mu'(P), \theta) \in G$$

or equivalently

$$\mu' \in [\![P]\!]_D.$$

Applying $v$ to the last relation we have that

$$v(\mu') \in v(\llbracket P \rrbracket_D)$$

and thus

$$\mu \in v(\llbracket P \rrbracket_D).$$

- Inductive step:
  - $P$ is $P_1$ $AND$ $P_2$.
    We have $v(\llbracket P_1 \rrbracket_D) = \llbracket P_1 \rrbracket_{v(D)}$ and $v(\llbracket P_2 \rrbracket_D) = \llbracket P_2 \rrbracket_{v(D)}$ from the inductive hypothesis. We will prove that $v(\llbracket P_1\ AND\ P_2 \rrbracket_D) = \llbracket P_1\ AND\ P_2 \rrbracket_{v(D)}$.

    Let $\mu \in v(\llbracket P_1\ AND\ P_2 \rrbracket_D)$. Therefore there exists a conditional mapping $\mu' = (\nu', \theta') \in \llbracket P_1\ AND\ P_2 \rrbracket_D$ such that $\mu = v(\mu')$ and $\mathbf{M}_\mathcal{L} \models v(\theta')$. Because $\llbracket P_1\ AND\ P_2 \rrbracket_D = \llbracket P_1 \rrbracket_D \bowtie \llbracket P_2 \rrbracket_D$, there exist possibly compatible conditional mappings $\mu_1' = (\nu_1', \theta_1')$ and $\mu_2' = (\nu_2', \theta_2')$ such that $\mu' = \mu_1' \bowtie \mu_2'$, $\mu_1' \in \llbracket P_1 \rrbracket_D$, and $\mu_2' \in \llbracket P_2 \rrbracket_D$. Because of Proposition E.1 and the fact that $\mathbf{M}_\mathcal{L} \models v(\theta')$, we have

    $$\mu = v(\mu') = v(\mu_1' \bowtie \mu_2') = v(\mu_1') \bowtie v(\mu_2').$$

    Since $\mathbf{M}_\mathcal{L} \models v(\theta')$ it also holds $\mathbf{M}_\mathcal{L} \models v(\theta_1')$ and $\mathbf{M}_\mathcal{L} \models v(\theta_2')$. Therefore, $v(\mu_1') \in v(\llbracket P_1 \rrbracket_D)$ and $v(\mu_2') \in v(\llbracket P_2 \rrbracket_D)$. Notice also that because $\mu_1'$ and $\mu_2'$ are possibly compatible, $v(\mu_1')$ and $v(\mu_2')$ are compatible. Therefore,

    $$v(\mu_1') \bowtie v(\mu_2') \in v(\llbracket P_1 \rrbracket_D) \bowtie v(\llbracket P_2 \rrbracket_D)$$

    which is equivalent to

    $$\mu \in v(\llbracket P_1 \rrbracket_D) \bowtie v(\llbracket P_2 \rrbracket_D).$$

    From the equalities of the inductive hypothesis, we now get

    $$\mu \in \left( \llbracket P_1 \rrbracket_{v(D)} \bowtie \llbracket P_2 \rrbracket_{v(D)} \right)$$

    which is equivalent to

    $$\mu \in \llbracket P_1\ AND\ P_2 \rrbracket_{v(D)}.$$

    This proof establishes that

    $$v(\llbracket P_1\ AND\ P_2 \rrbracket_D) \subseteq \llbracket P_1\ AND\ P_2 \rrbracket_{v(D)}.$$

    The other direction of the proof is similar and goes as follows.

    Let $\mu$ be a mapping such that $\mu \in \llbracket P_1\ AND\ P_2 \rrbracket_{v(D)}$. Then $\mu \in \left( \llbracket P_1 \rrbracket_{v(D)} \bowtie \llbracket P_2 \rrbracket_{v(D)} \right)$, which due to the inductive hypothesis gives us

    $$\mu \in v(\llbracket P_1 \rrbracket_D) \bowtie v(\llbracket P_2 \rrbracket_D).$$

    Therefore, there exist compatible mappings $\mu_1 \in v(\llbracket P_1 \rrbracket_D)$ and $\mu_2 \in v(\llbracket P_2 \rrbracket_D)$ such that $\mu = \mu_1 \bowtie \mu_2$. Thus, there exist conditional mappings $\mu_1' = (\nu_1', \theta_1') \in \llbracket P_1 \rrbracket_D$ and $\mu_2' = (\nu_2', \theta_2') \in \llbracket P_2 \rrbracket_D$ such that $\mu_1 = v(\mu_1')$, $\mu_2 = v(\mu_2')$, $\mathbf{M}_\mathcal{L} \models v(\theta_1')$ and $\mathbf{M}_\mathcal{L} \models v(\theta_2')$. Notice also that $\mu_1'$ and $\mu_2'$ are possibly compatible.
    From Proposition E.1 and the fact that $\mu = \mu_1 \bowtie \mu_2$, we have

    $$\mu = \mu_1 \bowtie \mu_2 = v(\mu_1') \bowtie v(\mu_2') = v(\mu_1' \bowtie \mu_2').$$

    Because $\mu_1' \in \llbracket P_1 \rrbracket_D$, $\mu_2' \in \llbracket P_2 \rrbracket_D$, and $\mu_1', \mu_2'$ are possibly compatible, we have

    $$\mu_1' \bowtie \mu_2' \in \llbracket P_1 \rrbracket_D \bowtie \llbracket P_2 \rrbracket_D.$$

    Now let $\mu' = (\nu', \theta')$ be a conditional mapping such that $\mu' = \mu_1' \bowtie \mu_2'$. Since $\mathbf{M}_\mathcal{L} \models v(\theta_1')$ and $\mathbf{M}_\mathcal{L} \models v(\theta_2')$, the definition of join of two compatible mappings gives us $\mathbf{M}_\mathcal{L} \models v(\theta')$. Therefore we can apply the valuation $v$ to $\mu'$ and get

    $$v(\mu') \in v(\llbracket P_1 \rrbracket_D \bowtie \llbracket P_2 \rrbracket_D).$$

    From this and the fact that $v(\mu') = v(\mu_1' \bowtie \mu_2') = \mu$ we get

    $$\mu \in v(\llbracket P_1\ AND\ P_2 \rrbracket_D).$$

  - $P$ is $P_1$ $UNION$ $P_2$.
    We have $v(\llbracket P_1 \rrbracket_D) = \llbracket P_1 \rrbracket_{v(D)}$ and $v(\llbracket P_2 \rrbracket_D) = \llbracket P_2 \rrbracket_{v(D)}$ from the inductive hypothesis. We will prove that

    $$v(\llbracket P_1\ UNION\ P_2 \rrbracket_D) = \llbracket P_1\ UNION\ P_2 \rrbracket_{v(D)}.$$

    A mapping $\mu$ is in $\llbracket P_1\ UNION\ P_2 \rrbracket_{v(D)}$ iff $\mu \in \llbracket P_1 \rrbracket_{v(D)} \cup \llbracket P_2 \rrbracket_{v(D)}$, which due to the inductive hypothesis is equivalent to $\mu \in v(\llbracket P_1 \rrbracket_D) \cup v(\llbracket P_2 \rrbracket_D)$, which can be seen to be equivalent to $\mu \in v(\llbracket P_1 \rrbracket_D \cup \llbracket P_2 \rrbracket_D)$, which is equivalent to

    $$\mu \in v(\llbracket P_1\ UNION\ P_2 \rrbracket_D).$$

  - $P$ is $P_1$ $FILTER$ $R$.
    We have $v(\llbracket P_1 \rrbracket_D) = \llbracket P_1 \rrbracket_{v(D)}$ from the inductive hypothesis. We will prove that

    $$v(\llbracket P_1\ FILTER\ R \rrbracket_D) = \llbracket P_1\ FILTER\ R \rrbracket_{v(D)}$$

    Without loss of generality, we give the proof only for the case of filters that are atomic $\mathcal{L}$-constraints (Definition 5.18). Let $\mu$ be in $\llbracket P_1\ FILTER\ R \rrbracket_{v(D)}$. By definition, this is equivalent to $\mu \in \llbracket P_1 \rrbracket_{v(D)}$ and $\mu \models R$. From the inductive hypothesis, we now have

    $$\mu \in v(\llbracket P_1 \rrbracket_D).$$

    Thus, there exists a conditional mapping $\mu' = (\nu', \theta') \in \llbracket P_1 \rrbracket_D$ such that $v(\mu') = \mu$ and $\mathbf{M}_\mathcal{L} \models v(\theta')$.
    Let now $\mu_1 = (\nu', \theta_1)$ be a conditional mapping with $\theta_1 = \theta' \wedge \nu'(R)$. Because $\mu \models R$, we have $\mathbf{M}_\mathcal{L} \models \mu(R)$. Therefore $\mathbf{M}_\mathcal{L} \models v(\nu'(R))$ since $v(\mu') = \mu$. Now notice that because $\mathbf{M}_\mathcal{L} \models v(\nu'(R))$ and $\mathbf{M}_\mathcal{L} \models v(\theta')$, we have $\mathbf{M}_\mathcal{L} \models v(\theta_1)$. Therefore $v(\mu_1)$ is well defined and we have $v(\mu_1) = v(\mu') = \mu$.
    The way $\mu_1$ and $\mu'$ have been defined above, together with the definition of the evaluation of FILTER graph patterns give us

    $$\mu_1 \in \llbracket P_1\ FILTER\ R \rrbracket_D.$$

    We can apply valuation $v$ to

    $$\mu_1 \in \llbracket P_1\ FILTER\ R \rrbracket_D$$

    and get

    $$v(\mu_1) \in v(\llbracket P_1\ FILTER\ R \rrbracket_D)$$

which is equivalent to $\mu \in v(\llbracket P_1 \ FILTER \ R \rrbracket_D)$.

This proof establishes that

$$\llbracket P_1 \ FILTER \ R \rrbracket_{v(D)} \subseteq v(\llbracket P_1 \ FILTER \ R \rrbracket_D).$$

The other direction of the proof is similar and goes as follows.

Let $\mu$ be a mapping in $v(\llbracket P_1 \ FILTER \ R \rrbracket_D)$. Then there exists a conditional mapping $\mu_1 = (\nu_1, \theta_1) \in \llbracket P_1 \ FILTER \ R \rrbracket_D$ such that $v(\mu_1) = \mu$ and $\mathbf{M}_\mathcal{L} \models v(\theta_1)$. Therefore, from the definition of FILTER evaluation there exists a conditional mapping $\mu_2 = (\nu_1, \theta_2)$ such that $\mu_2 \in \llbracket P_1 \rrbracket_D$, where $\theta_1 = \theta_2 \wedge \nu_1(R)$. Since $\mathbf{M}_\mathcal{L} \models v(\theta_1)$, then it holds that $\mathbf{M}_\mathcal{L} \models v(\theta_2)$ and $\mathbf{M}_\mathcal{L} \models v(\nu_1(R))$. Thus

$$v(\mu_2) = v(\mu_1) = \mu \in v(\llbracket P_1 \rrbracket_D).$$

Now using the inductive hypothesis, we have

$$\mu \in \llbracket P_1 \rrbracket_{v(D)}.$$

Because $\mathbf{M}_\mathcal{L} \models v(\nu_1(R))$ and $\mu = v(\mu_1)$, we have $\mathbf{M}_\mathcal{L} \models \mu(R)$. Thus we also have $\mu \models R$. Hence

$$\mu \in \llbracket P_1 \ FILTER \ R \rrbracket_{v(D)}.$$

□

THEOREM C.2. *The triple $\langle \mathcal{D}, Rep, \mathcal{Q}_{WD}^{C'} \rangle$ is a representation system.*

PROOF. The proof for Theorem C.2 is the same with the proof for Theorem C.1 and differs only in the inductive step for the OPTIONAL operator. Thus, in this case,

$$P \text{ is } P_1 \ OPT \ P_2.$$

We have $v(\llbracket P_1 \rrbracket_D) = \llbracket P_1 \rrbracket_{v(D)}$ and $v(\llbracket P_2 \rrbracket_D) = \llbracket P_2 \rrbracket_{v(D)}$ from the inductive hypothesis. We need to prove $v(\llbracket P_1 \ OPT \ P_2 \rrbracket_D) = \llbracket P_1 \ OPT \ P_2 \rrbracket_{v(D)}$ or equivalently

$$v(\llbracket P_1 \ OPT \ P_2 \rrbracket_D) = (\llbracket P_1 \rrbracket_{v(D)} \bowtie \llbracket P_2 \rrbracket_{v(D)}) \cup (\llbracket P_1 \rrbracket_{v(D)} \setminus \llbracket P_2 \rrbracket_{v(D)}). \quad (1)$$

Let $\mu \in v(\llbracket P_1 \ OPT \ P_2 \rrbracket_D)$. There exists conditional mapping $\mu' = (\nu', \theta') \in \llbracket P_1 \ OPT \ P_2 \rrbracket_D$ such that $\mu = v(\mu')$ and $\mathbf{M}_\mathcal{L} \models v(\theta')$. Since $\llbracket P_1 \ OPT \ P_2 \rrbracket_D = \llbracket P_1 \rrbracket_D \bowtie \llbracket P_2 \rrbracket_D = (\llbracket P_1 \rrbracket_D \bowtie \llbracket P_2 \rrbracket_D) \cup (\llbracket P_1 \rrbracket_D \setminus \llbracket P_2 \rrbracket_D)$,

$$\mu' \in (\llbracket P_1 \rrbracket_D \bowtie \llbracket P_2 \rrbracket_D) \text{ or } \mu' \in (\llbracket P_1 \rrbracket_D \setminus \llbracket P_2 \rrbracket_D).$$

For the first case, i.e., $\mu' \in (\llbracket P_1 \rrbracket_D \bowtie \llbracket P_2 \rrbracket_D)$ the proof is the same as in the proof of Theorem C.1, hence we finally get that

$$\mu \in (\llbracket P_1 \rrbracket_{v(D)} \bowtie \llbracket P_2 \rrbracket_{v(D)})$$

and thus from Formula (1) we have

$$\mu \in \llbracket P_1 \ OPT \ P_2 \rrbracket_{v(D)}.$$

For the second case, i.e., $\mu' \in (\llbracket P_1 \rrbracket_D \setminus \llbracket P_2 \rrbracket_D)$, and the definition of difference for sets of conditional mappings we distinguish two cases:

1. $\mu' \in \llbracket P_1 \rrbracket_D$ and for all $\mu_2 \in \llbracket P_2 \rrbracket_D$, $\mu'$ and $\mu_2$ are not compatible. Since $\mathbf{M}_\mathcal{L} \models v(\theta')$, we can apply valuation $v$ to $\mu'$ and get

$$\mu = v(\mu') \in v(\llbracket P_1 \rrbracket_D).$$

Since also every conditional mapping $\mu_2$ of $\llbracket P_2 \rrbracket_D$ is not compatible to $\mu'$ — and hence not possibly compatible to $\mu'$ — $v(\mu')$ is not compatible to every mapping $\mu'' \in v(\llbracket P_2 \rrbracket_D)$. Therefore,

$$v(\mu') \in (v(\llbracket P_1 \rrbracket_D) \setminus v(\llbracket P_2 \rrbracket_D))$$

which from our hypothesis is equivalent to

$$\mu \in (\llbracket P_1 \rrbracket_{v(D)} \setminus \llbracket P_2 \rrbracket_{v(D)}).$$

Hence, from Formula (1) we have

$$\mu \in \llbracket P_1 \ OPT \ P_2 \rrbracket_{v(D)}.$$

2. $\mu'$ is the conditional mapping $(\nu', \theta')$ and there exists conditional mapping $\mu'' = (\nu', \theta) \in \llbracket P_1 \rrbracket_D$ such that

   - $\mu''$ is not compatible to some mappings of $\llbracket P_2 \rrbracket_D$ and
   - for the rest mappings $\mu_i = (\nu_i, \theta_i) \in \llbracket P_2 \rrbracket_D$, $\mu''$ and $\mu_i$ are possibly compatible and

   $$\theta' \text{ is } \theta \wedge \left( \theta_i \supset \bigvee_x \neg(\mu''(x) \ EQ \ \mu_i(x)) \right)$$

   for $x \in dom(\mu'') \cap dom(\mu_i) \cap V_s$.

Since $\mathbf{M}_\mathcal{L} \models v(\theta')$, we can apply valuation $v$ to $\mu'$ and get

$$\mu = v(\mu') \in v(\llbracket P_1 \rrbracket_D \setminus \llbracket P_2 \rrbracket_D)$$

which can be written as

$$\mu = v(\mu') \in (v(\llbracket P_1 \rrbracket_D) \setminus v(\llbracket P_2 \rrbracket_D)).$$

To see this, notice that the above relation holds if and only if $v(\mu') \in v(\llbracket P_1 \rrbracket_D)$ and it is not compatible to every mapping $v(\mu_2)$ of $v(\llbracket P_2 \rrbracket_D)$. Since $\mathbf{M}_\mathcal{L} \models v(\theta')$ it holds $\mathbf{M}_\mathcal{L} \models v(\theta)$ and thus $v(\mu') = v(\mu'') \in v(\llbracket P_1 \rrbracket_D)$. Let us now take a mapping $\mu_2$ in $\llbracket P_2 \rrbracket_D$. Then, a) either $\mu''$, and consequently $\mu'$, is not compatible to $\mu_2$, or b) $\mu''$, and consequently $\mu'$, is possible compatible to $\mu_2$. For the first case $v(\mu')$ is also not compatible to $v(\mu_2)$. For the second case $v(\mu')$ is also not compatible to $v(\mu_2)$. To see this, notice that $v(\mu')$ and $v(\mu_2)$ become compatible only when $v(\mu'(x)) = v(\mu_2(x))$ for $x \in dom(\mu') \cap dom(\mu_2)$. In such cases, however, $\mathbf{M}_\mathcal{L} \not\models \theta'$ and thus $v(\mu') \notin v(\llbracket P_1 \rrbracket_D)$.

Continuing the proof, from our hypothesis, relation

$$\mu = v(\mu') \in (v(\llbracket P_1 \rrbracket_D) \setminus v(\llbracket P_2 \rrbracket_D))$$

now becomes

$$\mu \in (\llbracket P_1 \rrbracket_{v(D)} \setminus \llbracket P_2 \rrbracket_{v(D)})$$

and thus from Formula (1) we get

$$\mu \in \llbracket P_1 \ OPT \ P_2 \rrbracket_{v(D)}.$$

This proves that $v(\llbracket P_1 \ OPT \ P_2 \rrbracket_D) \subseteq \llbracket P_1 \ OPT \ P_2 \rrbracket_{v(D)}$. The other direction of the proof is similar. □

## D. PROOFS FOR SECTION 7

## D.1 Proof of Lemma 7.3

To show that $\bigcap Rep(D) = \bigcap Rep((D^{EQ})^*)$ we will first prove that

$$\bigcap Rep(D) \subseteq \bigcap Rep((D^{EQ})^*).$$

Let $t$ be an RDF triple such that $t \notin \bigcap Rep((D^{EQ})^*)$. Then, by definition of $Rep$ we get

$t \notin \bigcap \{H \mid$ there exists valuation $v$ such that
$$\mathbf{M}_\mathcal{L} \models v(\phi) \text{ and } H \supseteq v((G^{EQ})^*) \}.$$

Therefore, there exists valuation $v$ such that $\mathbf{M}_\mathcal{L} \models v(\phi)$ and $t \notin v((G^{EQ})^*)$, and thus

- either there is no conditional triple $(t', \theta') \in (G^{EQ})^*$ such that $\mathbf{M}_\mathcal{L} \models v(\theta')$, that is, $\mathbf{M}_\mathcal{L} \not\models v(\theta')$,
- or all conditional triples $(t', \theta') \in (G^{EQ})^*$ such that $\mathbf{M}_\mathcal{L} \models v(\theta')$ are such that $v(t') \neq t$.

Observe now that for conditional triples in $(G^{EQ})^*$, $\theta'$ can be written as $\bigvee_i \theta'_i$. So, if $(t', \theta') \in (G^{EQ})^*$, then $(t', \theta'_i) \in G^{EQ}$. Therefore, there will be a conditional triple $(t'', \theta'_i) \in G$, such that $t'$ and $t''$ possibly differ in their object position. In the following we construct $G$ and we show that $t \notin v(G)$ for this particular $v$.

For the first case above, and since $\mathbf{M}_\mathcal{L} \not\models v(\theta')$ we have that $\mathbf{M}_\mathcal{L} \not\models v(\theta'_i)$ for every $\theta'_i$, and thus such triples are dropped during application of valuation $v$ to $G$. Hence, if it was the case that $t \in \bigcap Rep(D)$, it would be so, only from the second case above.

Consider now the second case above and a triple $(t', \theta') \in (G^{EQ})^*$. Since $\mathbf{M}_\mathcal{L} \models v(\theta')$ and $(t', \theta'_i) \in G^{EQ}$, then some (or even all) $\theta'_i$ would be such that $\mathbf{M}_\mathcal{L} \models v(\theta'_i)$.

Let us now construct the conditional graph $G$ from $G^{EQ}$. Since $(t', \theta'_i) \in G^{EQ}$, then there exists conditional triple $(t'', \theta'_i) \in G$ such that $t'$ and $t''$ possibly differ in their object position. Let $t'$ be the e-triple $(s, p, o)$. Then:

1. If $o \in C$, then either $t''$ would be the same with $t'$, or it would have in its object position an e-literal $\_l$ such that $\phi \models \_l$ EQ $o$.
2. If $o \notin C$, then $t'$ and $t''$ would be the same.

Let us now apply valuation $v$ to $G$. Notice that $v(G)$ contains only RDF triples coming from conditional triples with a condition $\theta$ such that $v(\theta)$ is *true*. Thus, we could focus only on the conditional triples of $G$ with such conditions (it is clear from above, that such conditional triples do exist). To construct the RDF graph $v(G)$ it suffices to consider the two items above when applying $v$ to a conditional triple $(t'', \theta'_i)$ of $G$.

According to the second item and since $v(t') \neq t$ (see second case above), we have that $v(t'') \neq t$ as well. As for the first item, if $t' = t''$, then clearly we have $v(t'') \neq t$, since $v(t') \neq t$. Otherwise, $t''$ would be the triple $(s, p, \_l)$ such that $\phi \models \_l$ EQ $o$. In such a case, the application of $v$ to $t'$ would leave $t'$ unchanged, thus the RDF triple $t$ would contain in the object position a literal from $C$ and one that would be different from $o$ (this is because we are considering the second case for which $v(t') \neq t$). Since also $\phi \models \_l$ EQ $o$, then every valuation $v'$ that makes $v'(\phi)$ *true* it should make $v'(\_l$ EQ $o)$ *true* as well. Thus, such valuations would map the e-literal $\_l$ to the constant $o$. Since the valuation $v$ we consider is such a valuation, it maps $\_l$ to the constant $o$. Thus, again $v(t'') \neq t$.

Therefore, we showed that $v(G)$ cannot contain triple $t$ or equivalently that $t \notin v(G)$. Hence, from the definition of $Rep$ we have

$$t \notin \bigcap Rep(D)$$

which proves that

$$\bigcap Rep(D) \subseteq \bigcap Rep((D^{EQ})^*).$$

The other direction of the proof for showing

$$\bigcap Rep((D^{EQ})^*) \subseteq \bigcap Rep(D)$$

is similar.

## D.2 Proof of Theorem 7.4

Notice that the certain answer for $q$ over $D$ is the set

$$\bigcap [\![q]\!]_{Rep(D)}.$$

From the Representation Theorem and since $q \in \mathcal{Q}_{AUF}^{C'}$ it suffices to show that the algorithm computes the set

$$\bigcap Rep([\![q]\!]_D).$$

Notice that equation

$$\bigcap [\![q]\!]_{Rep(D)} = \bigcap Rep([\![q]\!]_D)$$

is a logical consequence of Definition 6.3 for the identity query.

Having Lemma 7.3 it now suffices to prove that the given algorithm computes the set

$$\bigcap Rep((([\![q]\!]_D)^{EQ})^*)$$

or, using the notation of Theorem 7.4, set

$$\bigcap Rep(((D_q)^{EQ})^*).$$

The first step of the algorithm evaluates $q$ over $D$, that is, it computes $D_q = (G_q, \phi)$, while the second step computes the EQ-completed form of $D_q$, that is, $(D_q)^{EQ}$, and then its normalized form, $((D_q)^{EQ})^*$.

It remains to show that step three computes exactly the intersection over the RDF graphs in $Rep(((D_q)^{EQ})^*)$.

Consider the set $\bigcap Rep(((D_q)^{EQ})^*)$ or equivalently the set

$\bigcap \{H \mid$ there exists valuation $v$ such that
$$\mathbf{M}_\mathcal{L} \models v(\phi) \text{ and } H \supseteq v(((D_q)^{EQ})^*) \}.$$

An RDF triple $t$ belongs to the above set iff for all valuations $v$ such that $\mathbf{M}_\mathcal{L} \models v(\phi)$, it holds

$$t \in v(((D_q)^{EQ})^*).$$

This is equivalent to requiring that a conditional triple $(t', \theta')$ exists in $H_q$ such that $\mathbf{M}_\mathcal{L} \models v(\theta')$ and $t = v(t')$ for all valuations $v$ such that $\mathbf{M}_\mathcal{L} \models v(\phi)$.

The first condition, that is, requiring that $\mathbf{M}_\mathcal{L} \models v(\theta')$ holds for all valuations $v$ such that $\mathbf{M}_\mathcal{L} \models v(\phi)$, is equivalent to requiring that $\phi \models \theta'$ holds, a requirement that step three imposes.

As for the second condition, equation $t = v(t')$ holds for any valuation $v$ such that $\mathbf{M}_\mathcal{L} \models v(\phi)$ iff $t'$ respects the following two cases:

- it does not contain any e-literal in the object position,
- it does contain an e-literal $\_l$ and all valuations $v$ above map $\_l$ to the same constant $c \in C$, which $t$ has it in its object position.

Since step three selects all conditional triples $(t', \theta)$ of $H_q$ such that $\phi \models \theta$ and $o \notin U$, the first case above is satisfied. The second case above is out of question: $H_q$ does not contain such a triple since all such e-literals have already been substituted by the respective constant $c \in C$ such that $\phi \models \_l$ EQ $c$.

Thus, step three computes exactly the set

$$\bigcap Rep(((D_q)^{\text{EQ}})^*).$$

## D.3   Proof of Theorem 7.6

Working similar to Proof D.2, it suffices to show that an RDF triple $t$ is in the certain answer of $q \in \mathcal{Q}_{AUF}^{C'}$ over $D$, that is, $t \in \bigcap Rep(\llbracket q \rrbracket_D)$, if and only if the following formula is valid:

$$(\forall \_l)(\phi(\_l) \supset \Theta(t, q, D, \_l)) \quad (1)$$

Let us now construct formula $\Theta(t, q, D, \_l)$ given the evaluation of $q$ over $D$, i.e., $\llbracket q \rrbracket_D = (G', \phi)$. Recall that formula $\Theta(t, q, D, \_l)$ is a disjunction of constraints $\theta_i$ for each conditional triple $(t'_i, \theta'_i) \in G'$ such that if $t$ and $t'_i$ have the same subject and predicate, $\theta_i$ is

- $\theta'_i$ if they agree in the object position as well,
- $\theta'_i \wedge (\_l$ EQ $o)$ if $t$ has the constant $o \in C$ at the object position and $t'_i$ has the e-literal $\_l \in U$ at the object position.

In every other case (i.e., $t$ and $t'_i$ do not agree in the subject and predicate or agree but the object of $t$ is not a constant from $C$ or the object of $t'_i$ is not an e-literal from $U$) no constraint $\theta_i$ is generated for those conditional triples and yet $\Theta(t, q, D, \_l)$ is taken to be $false$. Therefore, formula (1) is either unsatisfiable (we assume that the global constraint is always satisfiable) or of the form

$$(\forall \_l)(\phi(\_l) \supset \theta_1 \vee \ldots \vee \theta_k) \quad (2)$$

Consider now an RDF triple $t \notin \bigcap Rep(\llbracket q \rrbracket_D)$. Then there exists valuation $v$ such that $\mathbf{M}_\mathcal{L} \models v(\phi)$ and $t \notin v(G')$. Therefore, $G'$ contains conditional triples $(t', \theta')$ such that either

- $\mathbf{M}_\mathcal{L} \not\models v(\theta')$ or
- $\mathbf{M}_\mathcal{L} \models v(\theta')$ and $t \neq v(t')$.

Considering the first case above and since $\mathbf{M}_\mathcal{L} \not\models v(\theta')$, formula (2) would be unsatisfiable. To see this, notice that $\mathbf{M}_\mathcal{L} \not\models v(\theta')$ implies $\mathbf{M}_\mathcal{L} \not\models v(\theta'_i)$ and $\mathbf{M}_\mathcal{L} \not\models v(\theta'_i \wedge (\_l$ EQ $o))$ and thus the disjunction $\theta_i \vee \ldots \vee \theta_k$ in formula (2) is always $false$, and hence the whole formula is unsatisfiable.

To prove our result (i.e., that formula (1) is unsatisfiable for the specific RDF triple we considered), we have to show that formula (2) is unsatisfiable as well for the case in which $\mathbf{M}_\mathcal{L} \models v(\theta')$ and $t \neq v(t')$ (the second case above). Notice that $t \neq v(t')$ implies one of the following cases:

- $t$ and $t'$ do not agree in the subject or predicate position, or
- if they do, either they do not agree in the object position, or their objects are not of the proper kind (i.e., the object of $t$ is a constant from $C$ and the object of $t'$ is an e-literal from $U$), or if they are, then valuation $v$ does not map that e-literal to that constant.

From the first case, no constraint $\theta_i$ is included in formula (2). As for the second case, either no constraint $\theta_i$ is generated (the case in which they also differ in the object position) or $\theta_i$ is $\theta'_i \wedge (\_l$ EQ $o)$). As we pointed above, since valuation $v$ does not map the e-literal $\_l$ to the constant $o$, then $\theta_i$ is $false$. Hence, formula (2) is unsatisfiable as well.

The other direction of the proof is similar.

## D.4   Proofs for Section 7.1

PROPOSITION D.1.   *Let $D = (G, \phi)$ be an $RDF^i$ database, $q$ a query from $\mathcal{Q}_{AUF}^{C'}$ and $H$ an RDF graph. The certainty problem, $CERT_C(q, H, D)$, when the language of $\mathcal{L}$-constraints is ECL is coNP-complete.*

PROOF.   To decide $CERT_C(q, H, D)$, we have to check that $H \subseteq \bigcap \llbracket q \rrbracket_{Rep(D)}$ which from Definition 4.4 is equivalent to checking that $H \subseteq \llbracket q \rrbracket_{v(D)}$ for all valuations $v$ such that $\mathbf{M}_{ECL} \models v(\phi)$. Notice that the complement of this problem is to check whether there exists a valuation $v$ such that $\mathbf{M}_{ECL} \models v(\phi)$ and $H \not\subseteq \llbracket q \rrbracket_{v(D)}$. In other words, it suffices to find a valuation $v$ and a triple $t \in H$ such that $\mathbf{M}_{ECL} \models v(\phi)$ and $t \notin \llbracket q \rrbracket_{v(D)}$. This last problem is in NP, thus the certainty problem is in coNP.

Let us see why the complement problem defined above is in NP. We need only guess a valuation $v$ with length equal to the number of e-literals in $D$, check that $\mathbf{M}_{ECL} \models v(\phi)$, a computation that is in the AC complexity class, and then check that there exists $t \in H$ such that $t \notin \llbracket q \rrbracket_{v(D)}$. The steps for accomplishing the latter check, using Definition 4.6 for evaluating CONSTRUCT query forms of standard SPARQL [25], are the following:

1. Choose the next triple $t \in H$.
2. Loop over all candidate mappings $\mu$ for set $\llbracket P \rrbracket_{v(D)}$ generating a mapping per iteration, where $P$ is the graph patter of query $q$.
3. Check that $\mu \in \llbracket P \rrbracket_{v(D)}$.
4. Construct the renaming function $f_\mu$ based on the mapping $\mu$.
5. Generate set $S_\mu = \{\mu(f_\mu(E)) \cap (I \cup B) \times I \times T\}$.
6. Check whether $t \in S_\mu$. If yes, move to step 1, otherwise move to step 2. If there is no other mapping $\mu$ to check, return "yes". If there is no other triple to choose, return "no".

Step 2 above requires logarithmic space since the space required to store a candidate mapping $\mu$ from the set $\llbracket P \rrbracket_{v(D)}$ is $O(|P|(log|P| + log|D|))$ bits. This is because the mapping will contain $|P|$ variables and for each variable, it has to contain its value from $D$. The required space for each variable and value is $log|P|$ and $log|D|$, respectively. Since $q$ is fixed, the graph pattern $P$ is also fixed, therefore the space becomes logarithmic in the size of the database $D$.

Step 3 above can also be computed in LOGSPACE using the evaluation procedure EVAL presented in [26]. Further, since $q$ is fixed, then also the template $E$ and graph pattern $P$ are fixed. Thus, set $S_\mu$ of step 5 is of fixed size.

The coNP-hardness of $CERT_C(q, H, D)$ comes from a reduction from 3DNF tautology, which is known to be coNP-complete, and it is similar to the one employed in [6, Theorem 5.11, p. 118]. □

PROPOSITION D.2. *Let $D = (G, \phi)$ be an* RDF[i] *database, $q$ a query from $\mathcal{Q}^{C'}_{AUF}$ and $H$ an RDF graph. The certainty problem, $CERT_C(q, H, D)$, when the language of $\mathcal{L}$-constraints is one of* dePCL, diPCL, *or* RCL *is coNP-complete.*

PROOF. We sketch the proof for dePCL. The proof is similar for the cases of diPCL and RCL.

Similar to the proof for ECL, to decide $CERT_C(q, H, D)$, we have to check that $H \subseteq \bigcap [\![q]\!]_{Rep(D)}$ which from Definition 4.4 is equivalent to checking that $H \subseteq [\![q]\!]_{v(D)}$ for all valuations $v$ such that $\mathbf{M}_{dePCL} \models v(\phi)$. The complement of this problem is to check whether there exists a valuation $v$ such that $\mathbf{M}_{dePCL} \models v(\phi)$ and $H \not\subseteq [\![q]\!]_{v(D)}$. In other words, it suffices to find a valuation $v$ and a triple $t \in H$ such that $\mathbf{M}_{dePCL} \models v(\phi)$ and $t \notin [\![q]\!]_{v(D)}$. This last problem is in NP, thus the certainty problem is in coNP.

Let us see why the complement problem defined above is in NP. First, we use a non-deterministic Turing machine to guess in polynomial time a valuation $v$ that satisfies $\phi$ and then iterate over every triple $t$ of RDF graph $H$ checking whether $t \notin [\![q]\!]_{v(D)}$. This last check is done using the procedure described in the proof for ECL.

Let us see now how we can guess a valuation $v$ satisfying the global constraint $\phi$ of $D$ in polynomial time. To do this, we have to guess a rational number for every e-literal of the database $D$, substitute these values for e-literals in the global constraint $\phi$ and check that $\phi$ is *true* in polynomial time. Using Lemma 7.3 and Theorem 8.5 of [16], and Theorem 7.6 of our work, we can restrict the values over which the e-literals range only to a finite number of integers. The exact ranges are given in [16] and depend on the maximum absolute value of the constants appearing in formula $\phi$. Each value in these ranges takes up only polynomial amount of space with respect to the database size and the maximum absolute value of the constants of $\phi$, thus the guessing step can be done in polynomial time. Then, it is trivial to verify that $v(\phi)$ is *true*.

This proves that the complement of $CERT_C(q, H, D)$ is in NP and consequently that $CERT_C(q, H, D)$ is in coNP. The coNP-hardness of $CERT_C(q, H, D)$ follows from Proposition 3.1 of [35] where a sublanguage of dePCL/diPCL, similar to RCL, is considered that contains only the "less-than" predicate over rational or integer constants. Therefore, this lower bound holds for the languages diPCL and RCL as well. □

PROPOSITION D.3. *Let $D = (G, \phi)$ be an* RDF[i] *database, $q$ a query from $\mathcal{Q}^{C'}_{AUF}$ and $H$ an RDF graph. The certainty problem, $CERT_C(q, H, D)$, when the language of $\mathcal{L}$-constraints is* TCL *is in EXPTIME.*

PROOF. Since the satisfiability problem for conjunctions of TCL-constraints is known to be in PTIME [29], we can transform Formula (2) in DNF, construct a constraint network for each disjunct, and check them for consistency. This can be trivially solved in EXPTIME (DNF transformation). □

PROPOSITION D.4. *Let $D = (G, \phi)$ be an* RDF[i] *database, $q$ a query from $\mathcal{Q}^{C'}_{AUF}$ and $H$ an RDF graph. The certainty problem, $CERT_C(q, H, D)$, when the language of $\mathcal{L}$-constraints is* PCL *with the predicates of the RCC-5 calculus is in EXPTIME.*

PROOF. The above procedure applies also to the case of PCL restricted to the topological relations of RCC-5 when the involved constants are polygons in $V$-representation. In [20] it is shown that the satisfiability problem for such constraints can be decided in PTIME. □

# E. ADDITIONAL PROPOSITIONS

The next proposition shows that the result of applying a valuation to the join of two possibly compatible conditional mappings is the same as applying first the valuation to the conditional mappings and then computing their join as in standard RDF.

PROPOSITION E.1. *Let $v : U \to C$ be a valuation and $\mu_1 = (\nu_1, \theta_1)$, $\mu_2 = (\nu_2, \theta_2)$ be two possibly compatible conditional mappings such that $\mu_1 \bowtie \mu_2 = (\nu_3, \theta_3)$. Then*

$$v(\mu_1 \bowtie \mu_2) = v(\mu_1) \bowtie v(\mu_2)$$

*whenever these mappings are defined (i.e., whenever $\mathbf{M}_\mathcal{L} \models v(\theta_3)$ and therefore $\mathbf{M}_\mathcal{L} \models v(\theta_1)$ and $\mathbf{M}_\mathcal{L} \models v(\theta_2)$).*

The proof follows easily from the definition of join for conditional mappings and is omitted.

PROPOSITION E.2. *Let $D = (G, \phi)$ be an* RDF[i] *database, $q = (E, P)$ a CONSTRUCT query without blank nodes in $E$, and $v$ a valuation such that $\mathbf{M}_\mathcal{L} \models v(\phi)$. Then, $v([\![P]\!]_D) = [\![P]\!]_{v(D)}$ implies $v([\![q]\!]_D) = [\![q]\!]_{v(D)}$.*

PROOF. Let $[\![q]\!]_D$ be the RDF[i] database $D' = (G', \phi)$ where

$$G' = \bigcup_{\mu = (\nu, \theta) \in [\![P]\!]_D} \{(t, \theta) \mid t \in (\mu(f_\mu(E)) \cap ((I \cup B) \times I \times T))\}.$$

Then $v([\![q]\!]_D)$ is the RDF graph $v(D')$ where

$$v(D') = \bigcup_{\mu \in [\![P]\!]_D} \{v(t) \mid t \in (\mu(f_\mu(E)) \cap ((I \cup B) \times I \times T))$$
$$\text{and } \mu = (\nu, \theta) \text{ such that } \mathbf{M}_\mathcal{L} \models v(\theta)\}. \quad (1)$$

Likewise, let $[\![q]\!]_{v(D)}$ be the RDF graph $H$. According to the definition of the evaluation of CONSTRUCT queries on RDF graphs [25], $H$ is the following set:

$$H = \bigcup_{\mu \in [\![P]\!]_{v(D)}} \{\mu(f_\mu(E)) \cap ((I \cup B) \times I \times T)\} \quad (2)$$

To prove our proposition, we have to show that $H = v(D')$.

Let $t \in H$ be an RDF triple. Then, there exists a mapping $\mu \in [\![P]\!]_{v(D)}$ such that

$$t \in (\mu(f_\mu(E)) \cap ((I \cup B) \times I \times T)) \quad (3)$$

From our assumption that $v([\![P]\!]_D) = [\![P]\!]_{v(D)}$, we have

$$\mu \in v([\![P]\!]_D).$$

Therefore, there exists a conditional mapping $\mu' = (\nu', \theta') \in [\![P]\!]_D$ such that $\mathbf{M}_\mathcal{L} \models v(\theta')$ and $\mu = v(\mu')$. Since $\mu = v(\mu')$ relation (3) is written as

$$t \in \bigl(v(\mu'(f_\mu(E))) \cap ((I \cup B) \times I \times T)\bigr)$$

which is equivalent to the following

$$t \in v\bigl(\mu'(f_\mu(E)) \cap ((I \cup B) \times I \times T)\bigr). \quad (4)$$

From (1) and since $\mathbf{M}_\mathcal{L} \models v(\theta')$ and $\mu' \in [\![P]\!]_D$, we have

$$v\left(\mu'(f_\mu(E)) \cap ((I \cup B) \times I \times T)\right) \subseteq v(D').$$

Because also of relation (4) we get

$$t \in v(D').$$

Hence, we showed that every triple of $H$ is a triple of $v(D')$.

The other direction of the proof is similar and goes as follows.

Let $t \in v(D')$, then there exists conditional mapping $\mu = (\nu, \theta) \in [\![P]\!]_D$ and conditional triple $t_c = (t', \theta) \in G'$ such that $\mathbf{M}_\mathcal{L} \models v(\theta)$, $v(t_c) = v(t') = t$. From (1) we then have

$$t' \in (\mu(f_\mu(E)) \cap ((I \cup B) \times I \times T)). \quad (5)$$

Since $\mathbf{M}_\mathcal{L} \models v(\theta)$, $v(\mu)$ is defined and thus we have

$$v(\mu) \in v([\![P]\!]_D)$$

which from our assumption that $v([\![P]\!]_D) = [\![P]\!]_{v(D)}$ we get

$$\mu' = v(\mu) \in [\![P]\!]_{v(D)}.$$

Thus, applying valuation $v$ to (5) we get

$$v(t') \in v\left(\mu(f_\mu(E)) \cap ((I \cup B) \times I \times T)\right)$$

which is equivalent to

$$t \in (v(\mu(f_\mu(E))) \cap ((I \cup B) \times I \times T)).$$

Since $\mu' = v(\mu)$, the above relation becomes

$$t \in \left(\mu'(f_\mu(E)) \cap ((I \cup B) \times I \times T)\right).$$

From the above and because of (2) and $\mu' \in [\![P]\!]_{v(D)}$ we have

$$t \in \left(\mu'(f_\mu(E)) \cap ((I \cup B) \times I \times T)\right) \subseteq H$$

and hence

$$t \in H.$$

□